\documentclass[3p]{elsarticle}

\usepackage{amssymb}
\usepackage{amsmath,amssymb,amsfonts}
\usepackage{algorithmic}
\usepackage{graphicx}
\usepackage{textcomp}
\usepackage{xcolor}
\usepackage{framed}
\usepackage{listings}
\usepackage{xcolor}
\usepackage{graphicx}
\usepackage{caption}
\usepackage{amsmath}
\usepackage{amsfonts} 
\usepackage{CJK}
\usepackage{hyperref}
\usepackage{environ}
\usepackage{tikz}
\usepackage{hyperref}
\usepackage{booktabs}
\usepackage{multicol}
\usepackage{multirow}
\usepackage{threeparttable}
\usepackage{comment}
\usepackage{colortbl}
\usepackage[utf8]{inputenc}
\usepackage[linesnumbered,ruled,vlined]{algorithm2e}
\usepackage{textcomp}
\usepackage{pifont}
\usepackage{lineno}
\modulolinenumbers[5]
\definecolor{c1}{HTML}{7e0f12}
\NewEnviron{elaboration}{
\par
\begin{tikzpicture}
\node[rectangle,minimum width = 0.97\textwidth] (m) {\begin{minipage}{0.97\textwidth}\BODY\end{minipage}};
\draw[dashed] (m.south west) rectangle (m.north east);
\end{tikzpicture}
}

\renewcommand{\arraystretch}{1.5}

\journal{Information and Software Technology}

\begin{document}

\begin{frontmatter}

\begin{titlepage}
\begin{center}
\vspace*{1cm}

\textbf{\large Are Your Apps Accessible? A GCN-based Accessibility Checker for Low Vision Users}

\vspace{1.5cm}

Mengxi Zhang$^{a,b}$ (zmx19@mails.jlu.edu.cn), Huaxiao Liu$^{a,b}$ (liuhuaxiao@jlu.edu.cn), Shenning Song$^{a,b} (sunning2118@gmail.com), $Chunyang Chen$^{c}$ (chun-yang.chen@tum.de), Pei Huang$^{d}$ (huangpei@stanford.edu), Jian Zhao$^{e}$ (zhaojian@ccu.edu.cn) \\

\hspace{10pt}

\begin{flushleft}
\small  
$^a$ College of Computer Science and Technology, Changchun, Jilin University, 130012, Jilin, China \\
$^b$ Key Laboratory of Symbolic Computation and Knowledge Engineering of Ministry of Education, Changchun, Jilin University, 130012, Jilin, China \\
$^c$ Department of Computer Science, Heilbronn, Technical University of Munich, 85354, Germany \\
$^d$ Department of Computer Science, California, Stanford University, 94305, U.S.A \\
$^e$ College of Computer Science, Changchun, Changchun University, 130012, Jilin China \\

\vspace{1cm}
\textbf{Corresponding Author:} \\
Huaxiao Liu \\
College of Computer Science and Technology, Changchun, Jilin University, 130012, Jilin, China \\
Email: liuhuaxiao@jlu.edu.cn

\end{flushleft}        
\end{center}
\end{titlepage}



\title{Are Your Apps Accessible? A GCN-based Accessibility Checker for Low Vision Users}


\author{Mengxi Zhang\textsuperscript{a,b}}
\author{Huaxiao Liu\textsuperscript{a,b}\corref{cor}}
\ead{liuhuaxiao@jlu.edu.cn}
\author{Shenning Song\textsuperscript{a,b}}
\author{Chunyang Chen\textsuperscript{c}}
\author{Pei Huang\textsuperscript{d}}
\author{Jian Zhao\textsuperscript{e}}
\cortext[cor]{Corresponding author.}

\affiliation{organization={College of Computer Science and Technology},
            city={Changchun},
            university={Jilin University,},
            postcode={130012}, 
            state={Jilin},
            country={China}}
\affiliation{organization={Key Laboratory of Symbolic Computation and Knowledge Engineering of Ministry of Education},
				 city={Changchun},
            university={Jilin University,},
            postcode={130012}, 
            state={Jilin},
            country={China}}
\affiliation{organization={Department of Computer Science},
            city={Heilbronn},
		 university={Technical University of Munich,},
		 postcode={85354},
            country={Germany}}
\affiliation{organization={Department of Computer Science},
		  city={California},
		  university={Stanford University,},
	       postcode={94305},
	       country={U.S.A}}
\affiliation{organization={College of Computer Science},
		  city={Changchun},
	       university={Changchun University,},
		  postcode={130012},
		  state={Jilin},
		  country={China}}
\corref{Email address: liuhuaxiao@jlu.edu.cn {Huaxiao Liu}}
\begin{abstract}
Context: Accessibility issues (e.g., small size and narrow interval) in mobile applications (apps) lead to obstacles for billions of low vision users in interacting with Graphical User Interfaces (GUIs). 
Although GUI accessibility scanning tools exist, most of them perform rule-based check relying on complex GUI hierarchies. 
This might make them detect invisible redundant information, cannot handle small deviations, omit similar components, and is hard to extend. 
Objective: In this paper, we propose a novel approach, named ALVIN (Accessibility Checker for Low Vision), which represents the GUI as a graph and adopts the Graph Convolutional Neural Networks (GCN) to label inaccessible components. 
Method: ALVIN removes invisible views to prevent detecting redundancy and uses annotations from low vision users to handle small deviations. 
Also, the GCN model could consider the relations between GUI components, connecting similar components and reducing the possibility of omission.
ALVIN only requires users to annotate the relevant dataset when detecting new kinds of issues. 
Results: Our experiments on 48 apps demonstrate the effectiveness of ALVIN, with precision of 83.5\%, recall of 78.9\%, and F1-score of 81.2\%, outperforming baseline methods. 
In RQ2, the usefulness is verified through 20 issues submitted to open-source apps. 
The RQ3 also illustrates the GCN model is better than other models.
Conclusion: To summarize, our proposed approach can effectively detect accessibility issues in GUIs for low vision users, thereby guiding developers in fixing them efficiently.
\end{abstract}




\end{frontmatter}


\section{Introduction}\label{sec: introduction}
With the rapid development of mobile information technology and the broader needs of users, humans are far more dependent on mobile devices than before~\citep{Chen2020UnblindYA}.
Until 2022, Google Play and Apple Store have hosted approximately 2.65 million apps~\citep{GooglePlayNumber} and 3.79 million apps~\citep{appleStoreNumber} respectively, providing support for all aspects of people's daily life~\citep{Yan2019TheCS}.
Some of the longstanding ordinary activities, like shopping, social interaction, and entertainment have undergone radical changes.
However, as announced by the World Health Organization (WHO) in 2022~\citep{WorldHealthO}, there are nearly 2.2 billion visually impaired people worldwide, among which more than 70\% have low vision.
Such a defect prevents them from accessing information as easily as others~\citep{Oliveira2016AccessibilityOT, Silva2018ASO}.

Low vision is typically manifested in blurry or hazy vision.
As the National Eye Institute in America introduced, \emph{\textcolor{c1}{``low vision is a vision problem that makes it hard to do everyday activities. 
It can't be fixed with glasses, contact lenses, or other standard treatments like medicine or surgery''.}}~\footnote{\url{https://www.nei.nih.gov/learn-about-eye-health/eye-conditions-and-diseases/low-vision}}
Also, the best-corrected visual acuity measurement for low vision is between 20/200 to 20/70 (0.05 to 0.3 in China)~\citep{Xu2020PrevalenceAC}.
This visual defect makes these users only obtain vague and limited information when interacting with GUI~\citep{Kane2009FreedomTR, Ross2018ExaminingIB}.
It also makes it hard for low vision users to operate the apps effectively and enjoy their convenience.
With such issues and impacts, the development of GUIs should consider the experiences of low vision users, and perform effective accessibility testing \citep{Ross2020AnEL,Wei2016TamingAF,Bigham2010AccessibilityBD,Mack2021WhatDW}.

\begin{figure*}
\centering
\includegraphics[width=16.5cm]{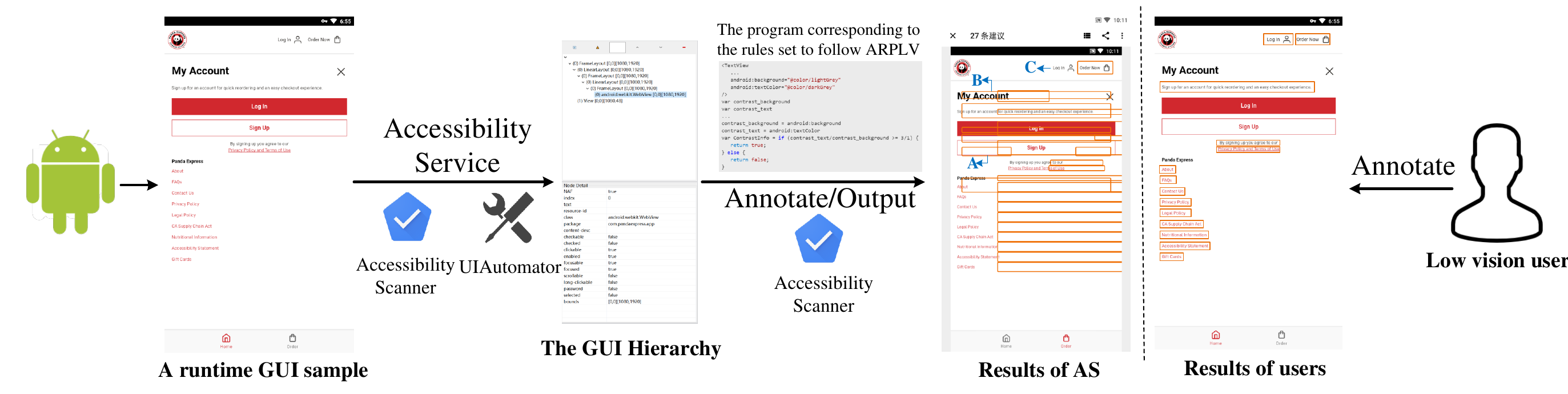}
\caption{The process of Accessibility Scanner to detect the accessibility issues in GUIs}
\label{fig: asexample}
\end{figure*}
To achieve this goal, the W3C drafted and released Accessibility Requirements for People with Low Vision (ARPLV) in 2016~\citep{LowVision}.
It illustrates the guidelines for the aspects of brightness and color, perceiving, spacing for reading, identifying elements, and so forth.
Also, the industry and academic community have proposed several practical tools, such as the Accessibility Scanner (AS)~\citep{AccessibilityScanner}, Mobile Accessibility Checker (MAC)~\citep{MobileAccessibilityChecker}, the Programmable UI mobile-automation (PUMA)~\citep{Hao2014PUMAPU}, the Mobile Automated Test Framework (MATE)~\citep{Eler2018AutomatedAT}, and Xbot~\citep{Chen2021AccessibleON}, which support capturing the inaccessible components in the GUIs automatically.

Figure~\ref{fig: asexample} shows how the Accessibility Scanner~\citep{AccessibilityScanner} detects accessibility issues in a runtime GUI sample. 
As we can see, when this tool starts performing detection, Android provides an Application Programming Interface (API) named Accessibility Service~\citep{AccessibilityService} to support it in obtaining the GUI hierarchies and attributes.
It is also consistent with the XML file parsed by uiautomator~\citep{UIautomator}~\footnote{UIAutomator is a GUI testing framework that is suitable for cross-app functional UI testing across system and installed apps.}.
Then, following the needs of ARPLV, it formulates various rules and applies them to the GUI hierarchies in the form of programs. 
Such programs enable this tool to detect whether there are accessibility issues in GUIs, along with what kind of issues they have.
The result regarding our presented GUI sample is shown in Figure~\ref{fig: asexample} (Results of AS), as well as the annotated result by low vision users is shown in Figure~\ref{fig: asexample} (Results of users). 
However, in the process of such detection, the GUI hierarchies are often diversified and nested, and the formulated rules are also hard to guarantee that they could be perfectly applied to various GUI hierarchies. 
Therefore, from a practical performance perspective, there are four kinds of problems with existing tools: \emph{(1) invisible redundant information is detected, (2) small deviations are not handled properly, (3) similar components might be omitted, and (4) it is hard to extend further.}
In more detail, we explain the four kinds of problems as follows.
\vspace{-0.42em}

\begin{enumerate}
\item \textbf{\emph{Invisible redundant information is detected.}} There are invisible views in the GUI hierarchies, such as \emph{$<$recyclerview$>$},\emph{$<$drawerlayout$>$}, and \emph{$<$viewpager$>$}, and existing tools would also detect these views, leading to redundant information in the results.
As we can see the invisible component $A$ in Figure~\ref{fig: asexample} (Results of AS), which has been considered to be problematic by AS.
However, low vision users do not annotate it because they think it is just a blank area.
\item \textbf{\emph{Small deviations are not handled properly.}} This problem refers to the fact that existing tools would annotate components with only small deviations from the ARPLV to be inaccessible, because of the enforced constraints of the rules. 
Component $B$ in Figure~\ref{fig: asexample} (Results of AS) shows such a kind of problem. 
The width of this component is $23px$, which is only $1px$ away from the $24px$ required by the standard, while AS marks it to be inaccessible.
In fact, such components with small deviations could be seen by low vision users, and are not annotated in Figure~\ref{fig: asexample} (Results of users).
\item \textbf{\emph{Similar components might be omitted.}} In the navigation bar of GUIs or components with a similar structure, existing tools may ignore components with accessibility issues.
As component $C$ in Figure~\ref{fig: asexample} (Results of AS) shows, it has a similar structure to the other component, but it is not marked. 
At the same time, low vision users deem that these two components should be regarded as inaccessible and annotate them in Figure~\ref{fig: asexample} (Results of users), because they cannot see both of them clearly.
\item \textbf{\emph{It is hard to extend further.}} Existing tools are normally developed based on rules, while the process of manually defining rules is extremely tedious, and it is hard to extend further.
Considering the accessibility issue of unclear alert information that none of the existing tools can detect.
If these tools want to detect this kind of issue, developers need to analyze large-scale GUIs, conduct user surveys, and summarize possible solutions.
Meanwhile, these solutions require multiple rounds of testing and tuning before they can be integrated into the tool and released.
This process will undoubtedly consume a lot of manpower and time, hindering development progress.
\end{enumerate}

\vspace{-0.42em}
In this work, we propose a tool, named ALVIN (\textbf{A}ccessibility Checker for \textbf{L}ow \textbf{Vi}sio\textbf{n}), to check the accessibility issues in Android apps.
Building upon our prior work~\citep{10338828}, we identified common accessibility issues in GUIs for users with low vision, including small sizes, narrow intervals, low color contrast, and unclear alert information, as well as the current status of accessibility (with 78.95\% of examined GUIs exhibiting accessibility issues in 500 apps). 
This provided the necessary background motivation for ALVIN and clarified which accessibility issues it should address. 
Thus, the method in this work is designed to check these four categories of accessibility issues commonly encountered by low vision users.
We further elucidate the fourth category of issues as follows: when users input incorrect or omitted information, the app provides alert information to remind users, however, these alerts are unclear or challenging to comprehend.
We then, formulate specific schemes to transform the visible components (e.g., image, button, and text), the mutually exclusive containers in GUIs, and their relations into graph structures, called GUI-graph.
The GUI-graph could ensure the overall structural integrity of GUIs as much as possible, while reducing the impact of redundant information on checking accessibility issues \textbf{(\emph{Solution for the problem} (1))}.
Subsequently, we apply the multiple classification model with Graph Convolutional Neural Network (GCN)~\citep{Kipf2017SemiSupervisedCW} for capturing the accessibility issues in the GUIs.
In this regard, we annotate 2,390 GUIs to train and validate our model (Section~\ref{sec: approach}) with the assistance of low vision users.
The annotation process begins with independent annotations by low vision users, followed by a collective discussion to determine a final annotation for each GUI component.
Such annotations are not bound by the rules, but are based on the practical experience of users, so that the small deviation can be handled well \textbf{(\emph{Solution for the problem} (2))}.
Also, the GCN model can fully consider the relationship between components, and components in the navigation bar or similar structure are usually in the same container and will be linked together. 
Therefore, if a component has accessibility issues, the other components connected to it will be affected, resulting in an increased probability of having accessibility issues \textbf{(\emph{Solution for the problem} (3))}.
Meanwhile, ALVIN only needs users to annotate the datasets when detecting new kinds of accessibility issues, without formulating cumbersome and complex rules \textbf{(\emph{Solution for the problem} (4))}.
We know that this is also labor-intensive, but much easier to perform than formulating rules.

We evaluate ALVIN based on 371 GUIs from 48 apps, showing that this tool could effectively capture accessibility issues in GUIs.
Also, the average performance of ALVIN outperforms the other rule-based baseline tools, consisting of the AS~\citep{AccessibilityScanner} released by Google in 2016, the MATE~\citep{Eler2018AutomatedAT} developed by Eler et al. to make mobile apps accessible, and the Xbot~\citep{Chen2021AccessibleON} proposed by Chen et al. to facilitate app accessibility testing and automatically collect accessibility issues.
Our conducted ablation experiments indicate that the feature attributes matrix, two convolutional layers, and the fully connected layer are all essential components of our method.
Furthermore, we explore the usefulness of ALVIN by submitting issues to 20 open-source apps, and encouragingly, 18 of 20 submitted issues are fixed or under fixing.
After conducting the aforementioned experiments, we also compare the performance of GCN model with Graph Attention Network (GAT)~\citep{Velickovic2017GraphAN}, Support Vector Machine (SVM)~\citep{Steinwart2008SupportVM}, Convolutional Neural Network (CNN)~\citep{LeCun2010ConvolutionalNA}, Random Forests (RF)~\citep{Breiman2001RandomF}, Multi-View Robust Graph Representation Learning (MGRL)~\citep{Ma2023MultiViewRG}, and Multi-Level Graph Relation Network (MuL-GRN)~\cite{Zhang2023MuLGRNMG}.
The results show that GCN model outperforms other baseline machine learning (ML) models, with the highest $Accuracy$ of 87.93\%.
These findings indicate that the effectiveness and usefulness of our proposed ALVIN are good enough for practical use (Section~\ref{sec: evaluation}). 

In summary, this paper makes the following three contributions:
\begin{itemize}
\item This is the first study on capturing the accessibility issues in Android apps by deep learning method with the cooperation of low vision users.
\item Our proposed tool, ALVIN, can fully consider the relations between components, as well as avoid invisible redundant annotations caused by the constrained rules.
\item We open the source of our reusable datasets consisting of available apps and labeled GUIs~\citep{DataCode}, which support the community conducting further research.
\end{itemize}

The remainder of this paper is organized as follows:
Section~\ref{sec: related work} reviews many works closely related to ALVIN, and motivates our work.
Section~\ref{sec: approach} presents in detail how the ALVIN is designed.
Section~\ref{sec: evaluation} evaluates the effectiveness and usefulness of ALVIN and compares the GCN model with other models.
This section also illustrates the threats to validity of ALVIN.
Section~\ref{sec: conclusion} draws the conclusion of this work.

\section{Related Work}\label{sec: related work}
In this section, we review works closely related to our approach, which falls into three categories: GUI understanding and GUI grouping, research on GUI accessibility, and GUI accessibility test.
From these three categories of works, we delve into the current research efforts in the direction of GUI accessibility and further analyze the limitations that serve to motivate our work.
\vspace{-0.42em}

\subsection{GUI understanding and GUI grouping}
The upstream work related to the understanding and grouping of GUIs can provide a solid foundation for accessibility testing of GUIs. 
This is also a primary avenue for researchers to explore various types of information within GUIs. 

From the perspective of GUI understanding, Liu et al~\cite{Liu2020OwlES} developed OwlEye to analyze and understand UI display issues (e.g., text overlap, missing image, blurred screen) via visual understanding.
Also, Liu et al.~\cite{Liu2022NighthawkFA} proposed Nighthawk, which aimed to understand the elements of GUI from a visual perspective, and detected and located visual inconsistencies in GUI design.
Zhang et al.~\cite{Zhang2022UniRLTestUP} shifted the focus of understanding towards the GUI images themselves, proposing UniRLTest. 
This approach utilizes CV (Computer Vision) techniques to capture all widgets within GUI images and constructs a widget tree for each GUI, thereby facilitating independent testing of the GUI.
To group the fragmented elements in UI designs, Chen et al.~\cite{Chen2023EGFEEG} presented EGFE, an approach that leverages multimodal feature representation and a Transformer encoder to group fragmented UI elements. 
It addresses the detection and grouping of tiny GUI components, achieving high-quality front-end code generation.
Further, Xie et al.~\cite{Xie2022PsychologicallyinspiredUI} explored the use of Gestalt psychology principles for perceptual grouping of GUI widgets. 
It identifies visual patterns and groups related elements based on the laws of connectedness, similarity, proximity, and continuity.

The aforementioned strategies of understanding GUI and grouping GUI elements enhance the perceptual clarity of GUI designs. 
They also provide a more reasonable analytical approach for subsequent accessibility testing.

\subsection{Research on GUI accessibility}

The accessibility of GUIs determines whether visually impaired users can interact effectively with them.
However, existing apps often face challenges in addressing accessibility due to constraints in their requirements and developers' oversight. 
This lack of consideration for accessibility issues poses significant difficulties for visually impaired users. 
Current research systematically examined and analyzed the types and quantity of accessibility issues within apps, highlighting the need for developers to incorporate accessibility considerations into GUI design to enhance the overall user experience. 

In this context, Alshayban et al.~\citep{Alshayban2020AccessibilityII} analyzed the prevalence of accessibility issues in over 1,000 Android apps, and found that almost all apps are riddled with accessibility issues, hindering their use by disabled people. 
A similar investigation was also conducted by Chen et al~\citep{Chen2021AccessibleON}, which found that there are a large number of accessibility issues in existing apps and their repair processes need to be strengthened.
Further, Rodrigues et al.~\citep{Rodrigues2020OpenCO}, from the perspectives of newcomers, novices, and experts, summarized 13 challenges (e.g., performing a specific touchscreen gesture is hard) faced by blind users in their daily lives using smartphones.
Research in this direction also includes the large-scale epidemiology-inspired study by Ross et al.~\citep{Ross2020AnEL}, which characterizes the current state of Android accessibility, suggests improvements to the app ecosystem, and demonstrates analysis techniques that can be applied in further app accessibility assessments.
Further, in our prior work~\cite{10338828}, we analyzed the common types of accessibility issues faced by low-vision users, providing relevant background support for this work. 
Our previous research focused more on the relationships between components and used a relationship prediction task to suggest how to adjust the components that can satisfy the needs of low-vision users.
In contrast, the study in this paper distinguishes itself from our previous research by relying on various attributes of GUI components to perform a multi-classification task aimed at detecting and identifying accessibility issues within GUIs.
Overall, the objectives of the two approaches differ fundamentally, while previous work has provided significant data support and theoretical backing for the study in this paper.

In summary, researchers have conducted a series of empirical explorations, summarizing and discovering a significant number of accessibility issues within existing apps. 
Developers often overlook these problems due to constraints related to requirements and functionalities. 
This phenomenon underscores the necessity of an effective method to detect these issues, and further guide developers to fix them. 

\vspace{-0.42em}
\subsection{GUI accessibility test}
Accessibility testing, as a crucial branch of GUI testing, has witnessed the development of various tools and methods in both industry~\citep{AccessibilityScanner, MobileAccessibilityChecker, AAAT} and academia~\citep{Hao2014PUMAPU, Eler2018AutomatedAT, Zhang2018RobustAO, Oliveira2016AccessibilityOT, Hara2013CombiningCA, Zhang2021ScreenRC}.

The first work related to this research dates back to 1995, Richard et al.~\citep{Kline1995ImprovingGA} firstly presented UnWindows VI, a set of tools designed to assist low vision users of X Windows in effectively accomplishing two mundane yet critical interaction tasks.
Based on the image recognition method, Salehnamadi et al. \citep{Salehnamadi2021LatteUA} proposed a GUI accessibility checker named Latte by intercepting a large number of user-operated GUI instances and analyzing them with the Accessibility Service~\citep{AccessibilityService}.
Chen et al. \citep{Chen2021AccessibleON} proposed Xbot to check the accessibility issues in the GUIs, and made a better performance on collecting accessibility issues.
Besides, there are also many other tools, such as Google Accessibility Scanner \citep{AccessibilityScanner}, Mobile Accessibility Checker (MAC) \citep{MobileAccessibilityChecker}, PUMA \citep{Hao2014PUMAPU}, Automated Accessibility Test Tool (AATT) \citep{AAAT}, and MATE \citep{Eler2018AutomatedAT}.
After that, industry and academia have also proposed many tools after analyzing the GUIs and low vision users, such as component recommendation \citep{Long2020ADA}, sensitive widget identification \citep{Xiao2019IconIntentAI}, layout optimization \citep{Tomlinson2016TalkinAT}, and so forth.
To further improve the accessibility of apps, Chen et al. \citep{Chen2020UnblindYA} developed Labeldroid to automatically predict the labels of image-based buttons.   
Still, Labeldroid lacks considering the context of images, causing the labels generated to be incorrect.
To further improve the accuracy of predicted labels, Mehralian et al. \citep{Mehralian2021DatadrivenAR} proposed a more accurate label generation method utilizing the context information of images.
Also, Zhang et al.~\citep{Zhang2017InteractionPF} introduced interaction proxies as a strategy for runtime repair and enhancement of the accessibility of mobile applications.

In a comprehensive view, the aforementioned methods and tools offered detection methods for accessibility issues in GUI from various aspects. 
However, they share a limitation in that they are all rule-based. 
As illustrated in the specific examples analyzed in Section~\ref{sec: introduction}, rule-based methods often flag invisible redundant information, struggle with handling minor deviations effectively, overlook similar components, and face challenges in scalability. 
Therefore, there is an urgent need for a unified and accurate approach to detect GUI accessibility issues, providing more precise assistance to developers in identifying these problems. 
In essence, this is also a primary objective of our work.

\begin{figure*}
\centering
\includegraphics[width=14cm]{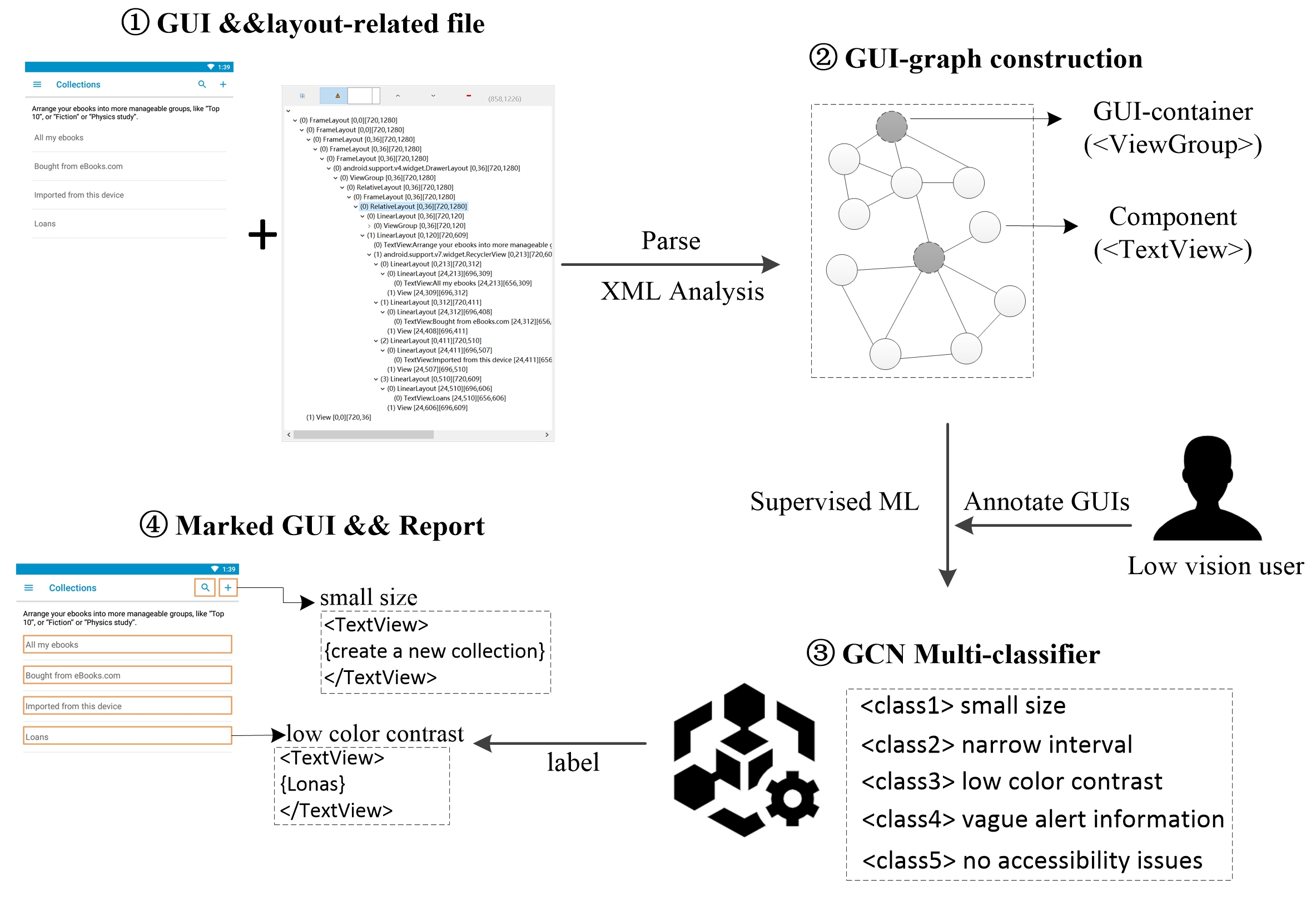}
\caption{The workflow of ALVIN.}
\label{fig: architecture}
\end{figure*}
\section{Methodology}\label{sec: approach}

The workflow of ALVIN (\textbf{A}ccessibility Checker for \textbf{L}ow \textbf{Vi}sio\textbf{n}) is shown in Figure~\ref{fig: architecture}.
First, given a GUI to be detected, we obtain its layout-related file by uiautomator \citep{UIautomator}.
Second, ALVIN converts the components and the structural information in the GUIs into the corresponding GUI-graphs via identifying nodes and determining edges between nodes. 
Then, with such GUI-graphs, ALVIN could conduct classification using the GCN multi-classifier, which is trained by the datasets annotated through low vision users (Section~\ref{sub: effectiveness}).
Finally, ALVIN marks the inaccessible components in GUIs with lighting wireframes, and generates reports to guide developers in fixing the issues.
As follows, we describe the ALVIN in detail.

\subsection{GUI-graph construction}
\begin{figure}
\centering
\includegraphics[width=15cm]{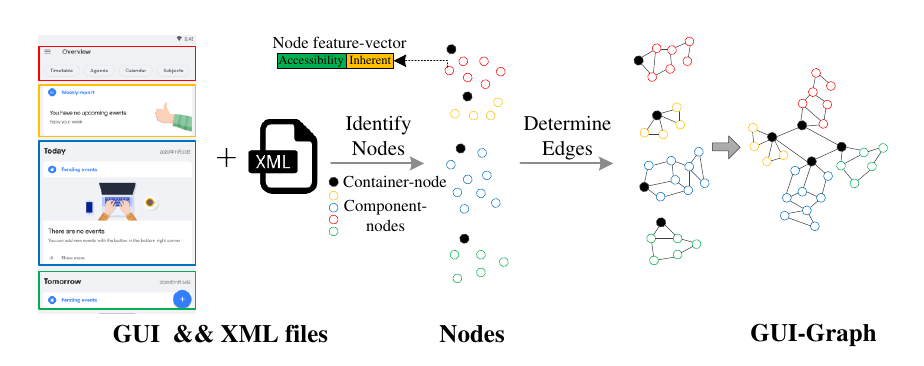}
\caption{The process of constructing GUI-graphs.}
\label{fig: construct}
\end{figure}

To achieve the goal of capturing the accessibility issues in the GUIs, we require a way to effectively represent GUIs.
Directly adopting the parsed results from uiautomator might be affected by the large amount of redundant information presented in its complex and diverse hierarchical structures.
Moreover, the GUIs might be treated as images, and make further operations using image processing methods~\citep{Liu2020OwlES, JaramilloAlczar2018AnAT}.
However, such a method would suffer many general concerns in locating the size and position of a component, as well as not being able to obtain obscure information about the component.
Therefore, refer to our previous work~\citep{10338828}, we construct the GUI-graphs for the components in the GUIs by fully considering the types of components, the positional relationships, and the attributes that each node contains.
Notably, for a GUI-graph, we mean a graph formed by connecting all components of a GUI according to their positional relationship, as opposed to a graph between different GUIs.
Then, following the process of Figure~\ref{fig: construct}, we explain how to convert a GUI and its XML file parsed by uiautomator into a GUI-graph.

\begin{figure}
\centering
\includegraphics[width=15cm]{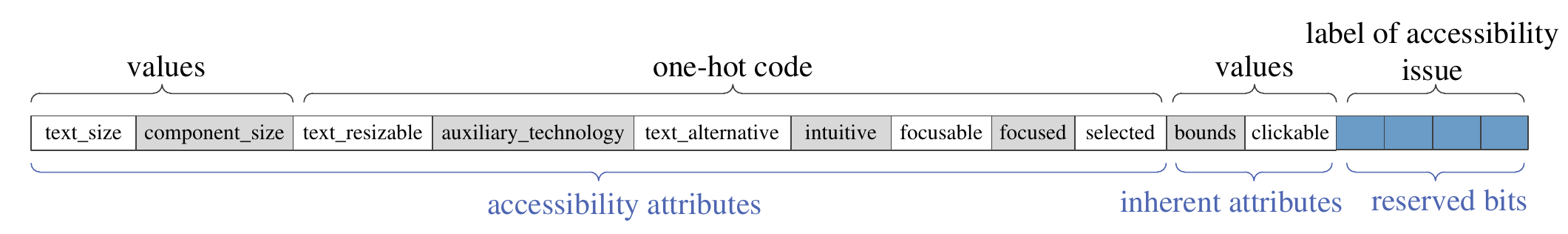}
\caption{The attributes of a component-node.}
\label{fig: attributes}
\end{figure}
\subsubsection{Node identification}\label{sub: nodes}

Identifying the nodes in the GUIs is the first step toward constructing the GUI-graphs.
There are two categories of nodes that need to be considered, involved in the component-nodes and the container-nodes.
The component-nodes are formed by views of \emph{$<$buttonview$>$}, \emph{$<$textview$>$}, \emph{$<$imageview$>$}, \emph{$<$listview$>$}, and \emph{$<$searchview$>$} in the GUIs, marked with unique ``resource\_id'' in the layout-related files.
For other views in these files, including \emph{$<$recyclerview$>$}, \emph{$<$drawerlayout$>$}, \emph{$<$viewpager$>$}, etc., we consider them invisible to low vision users and do not extract these views, since these users do not annotate them when checking GUIs.
Thus, our identification procedure can filter out the invisible redundant information, such as multi-component structures, format optimization widgets, hidden intermediate results, blank components and invisible BoxLayout.
The presence of such redundant information in the GUI can be attributed to two factors. 
Firstly, it may be the inclusion of blank information added by developers to ensure a consistent layout of other components within the GUIs. 
Secondly, it could be a result of intermediate computations during the GUI rendering process, such as layout calculations and graphical element rendering, which are stored as invisible information within the GUIs.
Then, the attributes of component-nodes are shown in Figure~\ref{fig: attributes}, which would be used for the subsequent GCN model (Section~\ref{sub: GCN}).
Specifically, the accessibility attributes in our work are ``text\_size'', ``component\_size'', ``text\_resizable'', ``auxiliary\_technology'', ``text\_alternative'', ``intuitive'', ``focusable'', ``focused'', and ``selected''.
The component-nodes also contain two inherent attributes that are ``bounds'' and ``clickable''.
Figure~\ref{fig: attributes} also presents the type and encoding (keep their values or one-hot code) of each attribute.
Notably, such aforementioned attributes are directly extracted or automatically calculated from the layout-related files parsed by uiautomator \citep{UIautomator} and AirTest \citep{AirTestIDE}.
The attributes that can be extracted directly consist of `text\_resizable'', ``auxiliary\_technology'', ``text\_alternative'', ``intuitive'', ``focusable'', ``focused'', ``selected'', ``bounds'', and ``clickable''.
While the attributes of ``text\_size'', ``component\_size'', and ``intuitive'' need to be calculated.
Using uiautomator, we can calculate the ``text\_size'' and ``component\_size'' from the difference between the coordinates of its vertices.
Version 1.2.6 of AirTest provides us with a way to compute the ``intuitive'', which is to calculate the difference between the RGB of the component and the RGB of the GUI background color.

Besides, the GUI layout files also provide 5 kinds of attributes, namely ``class'', ``package'', ``content-desc'', ``scrollable'', and ``password'', that we do not consider in this work.
As for the ``content-desc'', it is more inclined to discuss the issues encountered by blind users when using screen readers, not the main issue detected in our work.
Existing methods have been deeply explored on this issue, among which the method developed by Chen et al.~~\citep{Chen2020UnblindYA} to automatically generate the ``content-desc'' is the most prominent, and Mehralian et al.~\citep{Mehralian2021DatadrivenAR} have further optimized it.
While the ``password'' indicates whether the component is encrypted, ``scrollable'' refers to whether the component can scroll, along with ``class'' and ``package'' indicate its position in the source code.
These attributes cannot support us in capturing the accessibility issues in the components, so we ignore them.

Apart from the aforementioned attributes, our method aims to realize the classification of each component-node, so we need to save the four kinds of labels annotated by low vision users.
We achieve this goal by reserving bits in the attributes of nodes. 
One possible scheme is to reserve two bits to represent four kinds of labels, however, such coding has only four results, so it cannot represent the phenomenon that components are accessible.
Thus, we reserve four bits in the attributes of nodes, which represent the labels of four accessibility issues in each component-node, shown in the reserved bits in Figure~\ref{fig: attributes}. 
Where, starting from the left, the first reserved bit represents the issue of ``small size'', and when it is set to ``0'', it means this component does not have this kind of issue, while it has when it is set to ``1''. 
Similarly, the second bit refers to the issue of ``narrow interval'', the third bit is the issue of ``low color contrast'', and the fourth bit is the issue of ``unclear alert information''. 
When all bits are set to ``0'', it indicates that the node does not have any accessibility issues.
To train our subsequent model, low vision users annotate the components in GUIs, and we then map the annotations to the corresponding values of these four bits. 
While for predicting the accessibility issues in a new GUI, the reserved four bits in the node vector of each component are pre-set to ``null'' by default, since the categories for these nodes are unknown.

The other category of nodes is the container-nodes to represent the GUI-container.
Briefly, we only extract the uppermost \emph{$<$ViewGroup$>$} from the main branches of the XML tree as a container-node, to form the flattened layout instead of the complex nested layout.
The rationale is that, in the practical development of GUIs, components rarely nest more than 2 containers to avoid rendering speed degradation~\citep{Park2014TowardAM}.
With such a method, we can eliminate a large amount of redundant information as a result of uiautomator parse.
Notably, the container-nodes do not have any attribute.
It is only responsible for maintaining the basic structure of GUI-graphs.
Those nodes identified under the main branch will form a node group together to avoid a lot of redundancy when determining their edges subsequently.
Following the above process, we can obtain nodes capable of representing components and containers within the GUIs, seeing the \textbf{Nodes} in Figure~\ref{fig: construct}.

\subsubsection{Edges determination}
This step aims to determine the edges in the GUI-graphs. 
Based on the two categories of nodes and the node group defined above, there are three types of relationships among nodes.
They are, respectively, the edges between component-nodes, between component-nodes with container-nodes, and that among each container-nodes. 
Subsequently, we explain how to determine the above edges via using XML tree in detail.

Our method starts by determining the edges between component-nodes of the same node group in two ways.
One is that it traverses all nodes in the branches of XML tree according to the level order traversal method.
After that, our method determines the edge by analyzing whether two component-nodes are adjacent in the traversal sequence on the same or the adjacent level.
The other is that it would compare the position coordinates of each component-node and others in the same sub-graph, and then mark two nodes as adjacent nodes if their position coordinates are less than 5px (the default minimum interval between components in Android Developers Guideline for Layouts~\footnote{\url{https://developer.android.com/guide/topics/ui/declaring-layout}}, and it means that there could be no other components between these two components).
For the special structure components, like those that might be left-justified but vary greatly in length, it would be determined by whether these nodes are adjacent to their traversal sequence.
To sum up, if two component-nodes are marked as adjacent nodes, then there is an edge between them.

Following that, the second item aims to determine the edges connected between component-nodes and container-nodes within a node group.
With the structural information in the layout-related files, each container-node is served as the root of the corresponding sub-tree to connect the component-nodes.
Our method would maintain this connection.
Formally speaking, given a container-node $C$, the component-nodes contained in this container are $B_1, B_2,..., B_n$, and then the edges between $C$ and these component-nodes are $(C,B_1), (C,B_2),..., (C,B_n)$.

The last type is the edges between container-nodes. 
Our method connects the container-nodes from top to bottom, left to right in the GUI for each node group.
Further, it connects the last node group and the top node group to allow the information could propagate iteratively across this graph.
For instance, if the container-nodes in the position of GUI are sorted from top to bottom and left to right, then we obtain the sequence of $C_1, C_2, C_3,..., C_n$, while the edges connecting these container-nodes are $(C_1, C_2), (C_2, C_3),..., (C_{n-1}, C_n), (C_n, C_1)$.

Having drawn all edges, our attention turns to determining the weight of each edge, aiming to update it during the training process.
Inspired by ~\citep{Kipf2017SemiSupervisedCW}, we define the spatial order of neighboring nodes through the graph labeling process in the neighbor graph around the root.
Therefore, following formula~\ref{for: weight}, our method determines the weight between two nodes, and normalizes all of them with a linear method:
\begin{equation}
\label{for: weight}
w(c_i, c_j) = Out~of~degree (c_j)
\end{equation}
where $w(c_i, c_j)$ denotes the weight of the edge between node $c_i$ and $c_j$, while the $Out~of~degree (c_j)$ is the out of a degree of $c_j$.
In particular, the weights of edges between the container-nodes are set to 1 by default.
This is because such edges do not contain any feature information and are solely responsible for transferring information between node groups.
With all categories of edges being well drawn, we are able to derive a GUI-graph (like a \textbf{GUI-graph} in Figure~\ref{fig: construct}) to conduct the subsequent processing.

\subsection{GCN multi-classification Model} \label{sub: GCN}

\begin{figure*}
\centering
\includegraphics[width=15cm]{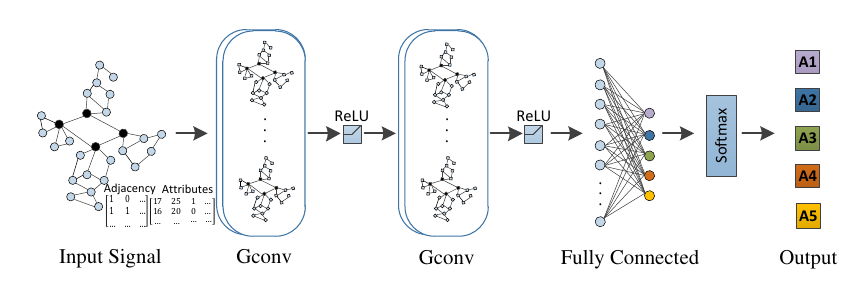}
\caption{The GCN multi-classification model.}
\label{fig: detecting}
\end{figure*}

In this part, we construct the GCN multi-classification model with the corresponding layers, and accordingly design a method to automatically check the inaccessible components in the GUIs.
Then, we utilize the annotated components by low vision users to provide this classification model with abundant corpora for training, validating, and testing.
Because the training set required for this model is annotated by low vision users, the classification results could align well with the experience of these users. 
Also, once there are new kinds of accessibility issues that need to be detected, our method only needs to annotate such issues in the GUIs, without formulating rules through testing, verification, iteration, and other complex means.
Compared with traditional rule-based methods, such a GCN-based method could bring the following benefits.
First, the relation between two components will affect each other, preventing one of them from being captured while the other from being omitted.
For example, in the navigation bars with similar components, if one component has accessibility issues, the probability that other components connected to it will have accessibility issues increases. 
This is because developers normally implement them in a uniform way.
Second, we avoid invalid components caused by cascading layout and neglecting to remove them, because low vision users do not include them in our datasets when annotating.
We recall that in Section~\ref{sub: nodes}, the GUI layout files contain invisible views for low vision users, like the \emph{$<$recyclerview$>$}, \emph{$<$drawerlayout$>$}, \emph{$<$viewpager$>$}, and so forth.
These views are not displayed in the GUI screens directly, so that low vision users could not see them, as well as would not annotate them.
Overall, the GUIs will be converted into GUI-graphs, and the multi-classification model will be trained together with the attributes of each node in these graphs.
Such a trained model can then achieve the goal of inputting a GUI-graph and the attributes of each component-node in this graph, then outputting whether each component has an accessibility issue and its type.
The brief procedure of this model is shown in Figure~\ref{fig: detecting}.

With the GUI-graphs being well constructed, we first convert them into the processable adjacency matrices, and further establish the Laplace matrices following their adjacency and degree matrices.
The adjacency matrices of GUI-graphs are square matrices, where their elements indicate whether pairs of vertices are adjacent or not in the graph~\citep{aylak2020WassersteinMF}.
For the Laplace matrices, they are calculated by subtracting the degree matrix from the adjacency matrix.
The degree matrix is a diagonal matrix that contains information about the degree of each vertex, that is, the number of edges attached to each vertex~\citep{Guo2020DynamicGC}.
Regarding these kinds of matrices, the adjacency matrix is typically used as an input to the model to represent the structure of the graph, while the Laplace matrix is responsible for eigen decomposition~\citep{Kipf2017SemiSupervisedCW}.
Notably, in such a process, all components (i.e., button, text, and image) in the GUIs would be regarded as component-nodes in the GUI-graphs, meanwhile, we connect each with the three categories of edge we define to reduce the impact of non-graph structures on model training, as well as the complex and diverse connectivity patterns.
As follows, we then design the network layers of our multiple classification model.
A total of four types of layers are presented in this structure, consisting of an input layer, a convolutional layer, a max-pooling layer, and an output layer with the Softmax function.
Among them, we apply the convolutional layer to extract features and share weights of the GUI-graph, as well as use the pooling layer to further compress them, and extract the main features~\citep{Kipf2017SemiSupervisedCW}.
With such two layers, the features in the data can be effectively optimized to reduce the output scale and the number of parameters required by the model.
This model starts with the input layer, which comprises the attributes matrix $A$ of the uni-directional GUI-graph and the adjacency matrix $X$ is composed of the edges between nodes.
Notably, the parameters of our models are consistent with the work~\citep{AbuElHaija2019NGCNMG}, since the basic architecture of our model is similar to it.
Notably, we have gone through a lot of debugging for the setting of each parameter and adopted the best solution.


After receiving the input signal of the matrices, the graphs would be passed into the convolutional layer.
This layer focuses on updating the current node based on its neighbor information until the balance is reached.
Meanwhile, we utilize the activation function (ReLU) to linearize the output after executing each convolutional layer. 
Such convolution and activation operations are performed twice in each network to ensure that the updated node information is sufficiently accurate.
The reason why we only design two-convolutional layers is that too many convolutional layers might cause the problem called ``degrade''~\citep{Wang2020ResGCNAMT}, it might lead each node to be categorized separately.
In detail, we conduct an experiment in Section~\ref{sub: ablation} to explore the performance of different times of convolution.
The results show that performing two operations yielded the best results in our test cases.
Soon after, we import the output of the first network to the max-pooling layer.
It aims at coarsening the output signal into sub-graphs so that node representation on coarsened graphs could get higher graph-level representations.
Then, we import the above pooling results into the second convolutional stage, and make further process.
On condition that all operations are well conducted, we send the pooling results to the fully connected layer, and determine the category for each node following the softmax function in the output layer.
Concerning such a process, we formalize the GCN multi-classification model as follows.
\begin{equation}
\begin{cases}
H^l = X & \text{$l~=~0$}\\
H^{l+1} = \sigma (\hat{D}^{\frac{1}{2}} \hat{A} \hat{D}^{\frac{1}{2}} H^l W^l) & \text{$l~\neq~0$}
\end{cases}
\label{for: formula2}
\end{equation}
where $X$ is the graph features, and the renormalized adjacency matrix $\hat{A} = \widetilde{D}^{\frac{1}{2}} \widetilde{A} \widetilde{D}^{\frac{1}{2}}$ is calculated in a pre-processing step.
This calculation might use the $A$ that refers to the adjacency matrix corresponding to the graph of GUI, and the identity matrix $I$.
The $\hat{D}$  is the degree matrix for each vertex of $\hat{A}$, and $\hat{D}^{\frac{1}{2}} \hat{A} \hat{D}^{\frac{1}{2}}$ aims to normalize $\hat{A}$, so as to prevent numerical instability in the training process.
Besides, we set $Y = \hat{D}^{\frac{1}{2}} \hat{A} \hat{D}^{\frac{1}{2}} H^l$, and the $H^l$ achieves the purpose of the spatial fitting.

As the training set is iterated with this model multiple times, we obtain a multi-classifier that can distinguish the components have which accessibility issues.
Then, to find out the most likely category of accessibility issues for each component, referenced with the prior study~\citep{Wang2020ResGCNAMT}~\citep{AbuElHaija2019NGCNMG}, we design the softmax function as a probability distribution function. 
Then, we take the largest posterior probability calculated by the following loss function (formula~\ref{for: loss}) as the final classification result. 
\begin{equation}
\label{for: loss}
loss = -\sum_{i=1}^ny_{i1}log\hat{y_{i1}} + y_{i2}log\hat{y_{i2}} + \cdots + y_{im}log\hat{y_{im}}
\end{equation}
where $n$ refers to the number of all samples, $m$ denotes the number of categories we defined, and $y_{i1}, y_{i2},.., y_{im}$ refer to the true distribution of the data, as well as $\hat{y_{i1}}, \hat{y_{i2}},.., \hat{y_{im}}$ are actual probability distribution obtained from the current training.
In general, our GCN model performs the node multi-classification task by taking the adjacency matrix, which represents the node relationships in the graph structure, and the attribute matrix, which represents the node features, as inputs. 
Subsequently, through the feature aggregation and forward propagation in the graph convolutional layer, the feature representations of each node are updated and new feature vectors are generated~\citep{Hu2019HierarchicalGC}. 
These updated features are then linearly transformed and mapped to the classification label space in the fully connected layer. 
Finally, a softmax layer is set as the output layer to transform each node's feature vector into a probability distribution representing the probabilities of belonging to each class.
This model therefore achieves the classification of each node, along with predicting different results for different components~\citep{Kipf2017SemiSupervisedCW}.

We now describe how our constructed GCN model predicts the accessibility issues of each component.
Given a GUI, there are multiple components and containers, which can be represented as a GUI-graph using ALVIN, and the node of this graph contains the vector of each component.
Then, such a graph is input into the trained-GCN model, and this model can predict the accessibility issues of each component according to the probability distribution of each node under different classes.
In practice, different GUIs contain different numbers of UI components, but our model requires that the dimensions of the GUI-graphs to be consistent when predicting the accessibility issues.
So, we define an exact threshold to expand the representation of GUI-graphs.
This is actually the method normally used by traditional ML models to convert variable length to fixed length when processing variable dimension inputs~\citep{Levin2013FixeddimensionalAE}.
In more detail, we count the number of UI components of each GUI we collected in Section~\ref{sub: effectiveness}.
Among them, a GUI contains a maximum of 37 components and a median of 19 components (as we filter out GUIs with 2 or fewer components).
Therefore, we set the threshold to 37 dimensions to ensure that the training sets and test sets can be processed successfully.
Notably, researchers can modify this threshold according to practical situations.
With this threshold, the GCN first expands the graph to 37 dimensions when inputting a GUI-graph, and subsequently performs the task of classification.
Such an extension essentially paddings ``0'' to the expanded dimension in the adjacency matrix, while the original features are unchanged. 
In this way, it will not only have no effect on the features of original graphs, but also make the model processes uniformly~\citep{Kittaneh2003BoundsFT}.

Notably, our method conducts a multi-classification task instead of a multi-labled classification, to determine which of the four possible accessibility issues may exist in a component or whether the component has no accessibility issues at all.
From a practical standpoint, GUI components may exhibit more than one accessibility issue.
Yet for low vision users, the issues they identify in the components are primarily determined by the problems that have the most significant impact on their operations, based on their subjective first impressions. 
For instance, consider a component with both the issues of small size and low color contrast, low vision users struggle to see the component primarily due to its low color contrast, while the size issue becomes less significant to them. 
This emphasizes the notion that a component with low color contrast remains imperceptible to users with low vision, regardless of its size. 
Furthermore, when low vision users annotate our provided dataset, their labeling is influenced by this subjective impression, leading them to mark the most primary accessibility issue they perceive, rather than multiple issues. 
Taking into account such distinctive user behavior perceptions and the resulting annotated dataset, our work is to conduct a multi-classification task that aims at detecting the most significant accessibility issue affecting low vision users for each component in GUIs. 
This task classifies components into five categories: the four accessibility issues and the component being accessible. 
Such detection results not only better reflect users' genuine perspectives but also provide valuable guidance to developers in effectively addressing the core issues.
In our work, the issue of small size is related to the length and width of components, the narrow interval is related to the vertex coordinates, the issue of low color contrast is associated with the value of RGB, and the vague alert information might be affected by all of this information. 
The final output of our model presents the five-tuple probability distribution of each node, denoted as $\left \langle p_1,p_2,p_3,p_4,p_5 \right \rangle$.
Where from $p_1$ to $p_5$, $p_1$ represents the probability of small size, $p_2$ is the probability of narrow interval, $p_3$ refers to the low color contrast, $p_4$ is the unclear alert information, as well as $p_5$ signifies the probability of the component being devoid of any accessibility issues.
The sum of the aforementioned five probability values is 1, and we select the maximum value among them to determine the category to which a component belongs.

The training, validation, and testing sets for this GCN model are mainly collected from the highly downloaded apps in Google Play, a total of 500 apps with 2,646 GUIs.
Then, we recruit 6 low vision users and 10 highly nearsighted volunteers to annotate the accessibility issues in these GUIs.
These users first annotate GUI components independently and then engage in a collective discussion to determine the final annotation for each component. 
This process ensures that there is only one label for each GUI component as perceived by low vision users. Additionally, the reason we do not directly allow low vision users to annotate through discussion is that without prior knowledge of the task, they would not know how to engage in meaningful discussions, making the annotation process difficult.
Afterward, we initially randomly selected 32 apps from the dataset and 16 other real-world apps in Google Play to create the test set. 
Then, out of the remaining 468 apps, 80\% are used for training, while the remaining 20\% are used for validation.
This selection ensures that the test set is independent of the training and validation sets.
More details about this collection, annotating, and evaluation process will be discussed in the data preparation of Section~\ref{sub: effectiveness}.

\begin{figure}
\centering
\includegraphics[width=10cm]{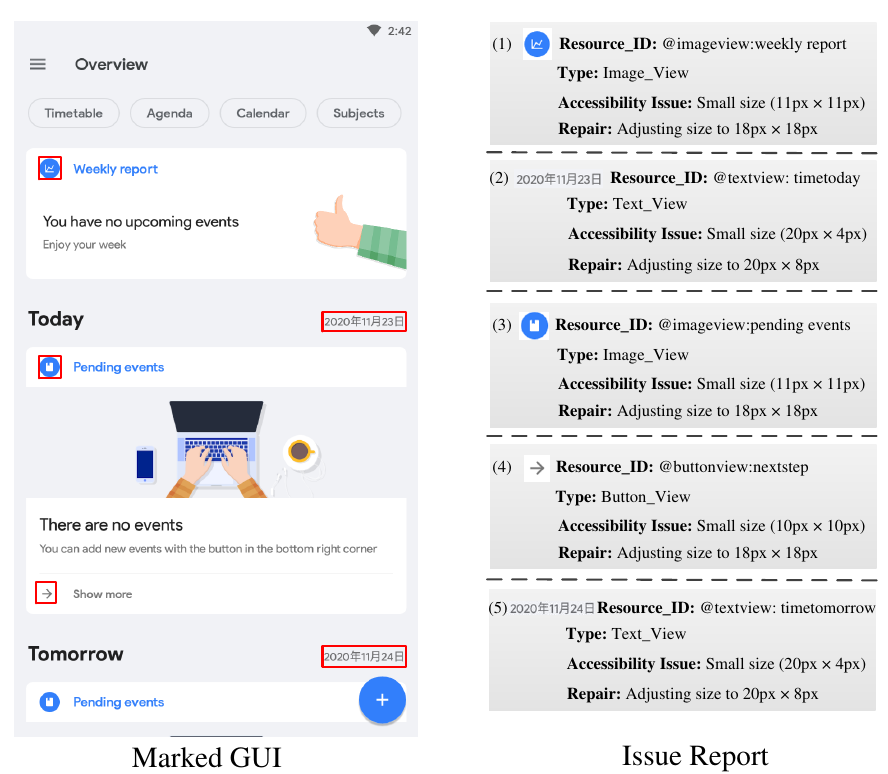}
\caption{An execution example of ALVIN.}
\label{fig: reports}
\end{figure}
\subsection{Implementation of ALVIN}\label{sub: implementation}
We implement ALVIN by packaging the GUI-graph construction module and multi-classification model as an executable file, to handle those GUIs that require to be checked.
The specific code and datasets have been published and uploaded to \emph{Zenodo}~\citep{DataCode}.

ALVIN could be executed with the following steps.
To begin, provide the layout-related files of GUIs and their corresponding screenshots.
Notably, if this GUI needs to input information, please complete it, to ensure that our tool could capture the accessibility issues accurately.
Then, ALVIN converts this GUI into its corresponding GUI-graph by parsing the XML file (in $\mathtt{GUI\_graph\_construct\_event()}$), and meanwhile, it could build the related node degree matrix and adjacency matrix of this GUI.
Subsequently, these matrices are fed into our constructed classifier (in $\mathtt{GCN(nn.Module)}$), and we obtain the classified component-nodes with specific accessibility issues.
ALVIN would mark the results in the GUI screenshots, and also generate the issues report following the strategies of the WCAG standard~\citep{LowVision}.

An execution example is shown in Figure~\ref{fig: reports}.
The left side is a GUI marked with lighting wireframes, indicating these components suffer accessibility issues, while the right side is the corresponding issues report, aiming to guide developers in fixing these issues.
The report presents the specific \emph{Resource\_ID}, \emph{Type}, \emph{Accessibility Issue}, and \emph{Recommended repair scheme} of each component with accessibility issues.
Accordingly, we summarize such reports could guide developers in the following three aspects.
One is to clearly and quickly assist developers in locating the components that need to be modified by the \emph{Resource\_ID} and \emph{Type}.
The second strength is that developers could understand exactly what kind of issue a component has, and assist developers in making modifications.
The last aspect is that these reports provide developers with modification strategies that they would use.

\vspace{-0.42em}
\section{Evaluation}\label{sec: evaluation}
To evaluate, and put in perspective, the effectiveness and usefulness of our proposed ALVIN, we conduct the relevant experiments. 
Based on the experimental results, we address the following three research questions:
\begin{itemize}
\item \textbf{RQ1 (effectiveness):} How effective is ALVIN in capturing accessibility issues?
\item \textbf{RQ2 (usefulness):} How useful is ALVIN in guiding developers to fix accessibility issues?
\item \textbf{RQ3 (model selection):} Why we select the GCN model to implemenet ALVIN?
\end{itemize}
\vspace{-0.42em}

\begin{table}\footnotesize
\renewcommand{\arraystretch}{1.2}
\tabcolsep=0.07cm
\caption{Effectiveness of ALVIN. Coms is the number of components. P is precision, R is recall, and F1 is F1-score.} \label{tab: effectiveness}
\begin{center}
\begin{tabular}{lcc|cccc|cccc|cccc|cccc}
\hline
\textbf{apps} & \textbf{GUIs} & \textbf{Coms} & \cellcolor{blue!2} & \cellcolor{blue!2} & \cellcolor{blue!2}\textbf{ALVIN} & \cellcolor{blue!2} & \cellcolor{green!2} & \cellcolor{green!2} & \cellcolor{green!2}\textbf{AS} & \cellcolor{green!2} & \cellcolor{purple!2} & \cellcolor{purple!2} & \cellcolor{purple!2}\textbf{MATE} & \cellcolor{purple!2} & \cellcolor{gray!2} & \cellcolor{gray!2} & \cellcolor{gray!2}\textbf{Xbot} & \cellcolor{gray!2} \cr
\cellcolor{gray!3}& \cellcolor{gray!3}& \cellcolor{gray!3} & \cellcolor{blue!2}TP & \cellcolor{blue!2}TN & \cellcolor{blue!2}FP & \cellcolor{blue!2}FN & \cellcolor{green!2}TP & \cellcolor{green!2}TN & \cellcolor{green!2}FP & \cellcolor{green!2}FN & \cellcolor{purple!2}TP & \cellcolor{purple!2}TN & \cellcolor{purple!2}FP & \cellcolor{purple!2}FN & \cellcolor{gray!2}TP & \cellcolor{gray!2}TN & \cellcolor{gray!2}FP & \cellcolor{gray!2}FN \cr
\hline
\cellcolor{gray!8}1. Zara & \cellcolor{gray!8}6  & \cellcolor{gray!8}42 & \cellcolor{blue!5}8 & \cellcolor{blue!5}27 & \cellcolor{blue!5}4 & \cellcolor{blue!5}3 & \cellcolor{green!5}11 & \cellcolor{green!5}27 & \cellcolor{green!5}4 & \cellcolor{green!5}1 & \cellcolor{purple!5}10 & \cellcolor{purple!5}21 & \cellcolor{purple!5}10 & \cellcolor{purple!5}3 & \cellcolor{gray!5}8 & \cellcolor{gray!5}23 & \cellcolor{gray!5}8 & \cellcolor{gray!5}11 \cr
\cellcolor{gray!3}2. Sky Map & \cellcolor{gray!3}5 & \cellcolor{gray!3}74 & \cellcolor{blue!2}22 & \cellcolor{blue!2}35 & \cellcolor{blue!2}7 & \cellcolor{blue!2}11 & \cellcolor{green!2}17 & \cellcolor{green!2}36 & \cellcolor{green!2}6 & \cellcolor{green!2}20 & \cellcolor{purple!2}17 & \cellcolor{purple!2}30 & \cellcolor{purple!2}12 & \cellcolor{purple!2}7 & \cellcolor{gray!2}30 & \cellcolor{gray!2}30 & \cellcolor{gray!2}12 & \cellcolor{gray!2}10 \cr
\cellcolor{gray!8}3. Duolinggo  & \cellcolor{gray!8}4 &\cellcolor{gray!8}48 & \cellcolor{blue!5}12 & \cellcolor{blue!5}25 & \cellcolor{blue!5}4 & \cellcolor{blue!5}9 & \cellcolor{green!5}10 & \cellcolor{green!5}24 & \cellcolor{green!5}5 & \cellcolor{green!5}16 & \cellcolor{purple!5}11 & \cellcolor{purple!5}19 & \cellcolor{purple!5}10 & \cellcolor{purple!5}10 & \cellcolor{gray!5}17 & \cellcolor{gray!5}21 & \cellcolor{gray!5}8 & \cellcolor{gray!5}11 \cr
\cellcolor{gray!3}4. Efla & \cellcolor{gray!3}7 & \cellcolor{gray!3}89 & \cellcolor{blue!2}32 & \cellcolor{blue!2}39 & \cellcolor{blue!2}6 & \cellcolor{blue!2}12 & \cellcolor{green!2}27 & \cellcolor{green!2}41 & \cellcolor{green!2}4 & \cellcolor{green!2}31 & \cellcolor{purple!2}27 & \cellcolor{purple!2}30 & \cellcolor{purple!2}15 & \cellcolor{purple!2}19 & \cellcolor{gray!2}32 & \cellcolor{gray!2}32 & \cellcolor{gray!2}13 & \cellcolor{gray!2}18 \cr
\cellcolor{gray!8}5. applebee's & \cellcolor{gray!8}8 & \cellcolor{gray!8}108 & \cellcolor{blue!5}54 & \cellcolor{blue!5}35 & \cellcolor{blue!5}9 & \cellcolor{blue!5}12 & \cellcolor{green!5}49 & \cellcolor{green!5}37 & \cellcolor{green!5}7 & \cellcolor{green!5}33 & \cellcolor{purple!5}49 & \cellcolor{purple!5}27 & \cellcolor{purple!5}17 & \cellcolor{purple!5}16 & \cellcolor{gray!5}53 & \cellcolor{gray!5}31 & \cellcolor{gray!5}13 & \cellcolor{gray!5}16 \cr
\cellcolor{gray!3}6. Fox & \cellcolor{gray!3}8 & \cellcolor{gray!3}124 & \cellcolor{blue!2}56 & \cellcolor{blue!2}48 & \cellcolor{blue!2}6 & \cellcolor{blue!2}14 & \cellcolor{green!2}51 & \cellcolor{green!2}47 & \cellcolor{green!2}7 & \cellcolor{green!2}30 & \cellcolor{purple!2}50 & \cellcolor{purple!2}36 & \cellcolor{purple!2}38 & \cellcolor{purple!2}3 & \cellcolor{gray!2}56 & \cellcolor{gray!2}39 & \cellcolor{gray!2}15 & \cellcolor{gray!2}21 \cr
\cellcolor{gray!8}7. Chipotle & \cellcolor{gray!8}6 & \cellcolor{gray!8}78 & \cellcolor{blue!5}20 & \cellcolor{blue!5}42 & \cellcolor{blue!5}4 & \cellcolor{blue!5}16 & \cellcolor{green!5}18 & \cellcolor{green!5}44 & \cellcolor{green!5}2 & \cellcolor{green!5}23 & \cellcolor{purple!5}12 & \cellcolor{purple!5}36 & \cellcolor{purple!5}10 & \cellcolor{purple!5}21 & \cellcolor{gray!5}18 & \cellcolor{gray!5}40 & \cellcolor{gray!5}6 & \cellcolor{gray!5}20 \cr
\cellcolor{gray!3}8. Snapseed & \cellcolor{gray!3}4 & \cellcolor{gray!3}25 & \cellcolor{blue!2}6 & \cellcolor{blue!2}14 & \cellcolor{blue!2}1 & \cellcolor{blue!2}4 & \cellcolor{green!2}7 & \cellcolor{green!2}14 & \cellcolor{green!2}1 & \cellcolor{green!2}6 & \cellcolor{purple!2}5 & \cellcolor{purple!2}9 & \cellcolor{purple!2}6 & \cellcolor{purple!2}6 & \cellcolor{gray!2}6 & \cellcolor{gray!2}11 & \cellcolor{gray!2}4 & \cellcolor{gray!2}13 \cr
\cellcolor{gray!8}9. Wise & \cellcolor{gray!8}9 & \cellcolor{gray!8}289 & \cellcolor{blue!5}92 & \cellcolor{blue!5}175 & \cellcolor{blue!5}8 & \cellcolor{blue!5}17 & \cellcolor{green!5}92 & \cellcolor{green!5}177 & \cellcolor{green!5}6 & \cellcolor{green!5}26 & \cellcolor{purple!5}75 & \cellcolor{purple!5}153 & \cellcolor{purple!5}30 & \cellcolor{purple!5}33 & \cellcolor{gray!5}90 & \cellcolor{gray!5}166 & \cellcolor{gray!5}17 & \cellcolor{gray!5}28 \cr
\cellcolor{gray!3}10. FSCK & \cellcolor{gray!3}6 & \cellcolor{gray!3}72 & \cellcolor{blue!2}29 & \cellcolor{blue!2}29 & \cellcolor{blue!2}5 & \cellcolor{blue!2}9 & \cellcolor{green!2}26 & \cellcolor{green!2}30 & \cellcolor{green!2}4 & \cellcolor{green!2}17 & \cellcolor{purple!2}27 & \cellcolor{purple!2}20 & \cellcolor{purple!2}14 & \cellcolor{purple!2}14 & \cellcolor{gray!2}27 & \cellcolor{gray!2}31 & \cellcolor{gray!2}3 & \cellcolor{gray!2}19 \cr
\cellcolor{gray!8}11. Waze & \cellcolor{gray!8}7 & \cellcolor{gray!8}80 & \cellcolor{blue!5}23 & \cellcolor{blue!5}40 & \cellcolor{blue!5}5 & \cellcolor{blue!5}13 & \cellcolor{green!5}18 & \cellcolor{green!5}41 & \cellcolor{green!5}4 & \cellcolor{green!5}25 & \cellcolor{purple!5}18 & \cellcolor{purple!5}30 & \cellcolor{purple!5}15 & \cellcolor{purple!5}17 & \cellcolor{gray!5}24 & \cellcolor{gray!5}29 & \cellcolor{gray!5}16 & \cellcolor{gray!5}18 \cr
\cellcolor{gray!3}12. MyChart & \cellcolor{gray!3}6 & \cellcolor{gray!3}64 & \cellcolor{blue!2}21 & \cellcolor{blue!2}31 & \cellcolor{blue!2}3 & \cellcolor{blue!2}9 & \cellcolor{green!2}18 & \cellcolor{green!2}30 & \cellcolor{green!2}4 & \cellcolor{green!2}16 & \cellcolor{purple!2}17 & \cellcolor{purple!2}24 & \cellcolor{purple!2}10 & \cellcolor{purple!2}13 & \cellcolor{gray!2}20 & \cellcolor{gray!2}27 & \cellcolor{gray!2}7 & \cellcolor{gray!2}16 \cr
\cellcolor{gray!8}13. Kardia & \cellcolor{gray!8}6 & \cellcolor{gray!8}53 & \cellcolor{blue!5}9 & \cellcolor{blue!5}29 & \cellcolor{blue!5}6 & \cellcolor{blue!5}9 & \cellcolor{green!5}13 & \cellcolor{green!5}32 & \cellcolor{green!5}3 & \cellcolor{green!5}8 & \cellcolor{purple!5}12 & \cellcolor{purple!5}21 & \cellcolor{purple!5}14 & \cellcolor{purple!5}6 & \cellcolor{gray!5}11 & \cellcolor{gray!5}30 & \cellcolor{gray!5}5 & \cellcolor{gray!5}14 \cr
\cellcolor{gray!3}14. RxSaver & \cellcolor{gray!3}8 & \cellcolor{gray!3}92 & \cellcolor{blue!2}32 & \cellcolor{blue!2}41 & \cellcolor{blue!2}7 & \cellcolor{blue!2}12 & \cellcolor{green!2}23 & \cellcolor{green!2}44 & \cellcolor{green!2}4 & \cellcolor{green!2}25 & \cellcolor{purple!2}26 & \cellcolor{purple!2}32 & \cellcolor{purple!2}16 & \cellcolor{purple!2}19 & \cellcolor{gray!2}30 & \cellcolor{gray!2}38 & \cellcolor{gray!2}10 & \cellcolor{gray!2}19 \cr
\cellcolor{gray!8}15. Wikipedia & \cellcolor{gray!8}5 & \cellcolor{gray!8}57 & \cellcolor{blue!5}27 & \cellcolor{blue!5}32 & \cellcolor{blue!5}4 & \cellcolor{blue!5}9 & \cellcolor{green!5}23 & \cellcolor{green!5}32 & \cellcolor{green!5}4 & \cellcolor{green!5}19 & \cellcolor{purple!5}24 & \cellcolor{purple!5}21 & \cellcolor{purple!5}15 & \cellcolor{purple!5}12 & \cellcolor{gray!5}25 & \cellcolor{gray!5}29 & \cellcolor{gray!5}7 & \cellcolor{gray!5}20 \cr
\cellcolor{gray!3}16. BBC & \cellcolor{gray!3}10 & \cellcolor{gray!3}109 & \cellcolor{blue!2}30 & \cellcolor{blue!2}70 & \cellcolor{blue!2}3 & \cellcolor{blue!2}6 & \cellcolor{green!2}29 & \cellcolor{green!2}71 & \cellcolor{green!2}2 & \cellcolor{green!2}9 & \cellcolor{purple!2}36 & \cellcolor{purple!2}65 & \cellcolor{purple!2}8 & \cellcolor{purple!2}0 & \cellcolor{gray!2}25 & \cellcolor{gray!2}68 & \cellcolor{gray!2}5 & \cellcolor{gray!2}16 \cr

\cellcolor{gray!8}17. Threads & \cellcolor{gray!8}8 & \cellcolor{gray!8}77 & \cellcolor{blue!5}29 & \cellcolor{blue!5}31 & \cellcolor{blue!5}6 & \cellcolor{blue!5}11 & \cellcolor{green!5}20 & \cellcolor{green!5}36 & \cellcolor{green!5}9 & \cellcolor{green!5}12 & \cellcolor{purple!5}15 & \cellcolor{purple!5}41 & \cellcolor{purple!5}12 & \cellcolor{purple!5}9 & \cellcolor{gray!5}28 & \cellcolor{gray!5}34 & \cellcolor{gray!5}2 & \cellcolor{gray!5}13 \cr
\cellcolor{gray!3}18. Twitch & \cellcolor{gray!3}7 & \cellcolor{gray!3}90 & \cellcolor{blue!2}21 & \cellcolor{blue!2}59 & \cellcolor{blue!2}5 & \cellcolor{blue!2}5 & \cellcolor{green!2}15 & \cellcolor{green!2}58 & \cellcolor{green!2}7 & \cellcolor{green!2}10 & \cellcolor{purple!2}10 & \cellcolor{purple!2}64 & \cellcolor{purple!2}8 & \cellcolor{purple!2}8 & \cellcolor{gray!2}20 & \cellcolor{gray!2}53 & \cellcolor{gray!2}1 & \cellcolor{gray!2}16 \cr
\cellcolor{gray!8}19. Revolut & \cellcolor{gray!8}7 & \cellcolor{gray!8}64 & \cellcolor{blue!5}18 & \cellcolor{blue!5}35 & \cellcolor{blue!5}4 & \cellcolor{blue!5}7 & \cellcolor{green!5}17 & \cellcolor{green!5}30 & \cellcolor{green!5}6 & \cellcolor{green!5}11 & \cellcolor{purple!5}13 & \cellcolor{purple!5}38 & \cellcolor{purple!5}7 & \cellcolor{purple!5}6 & \cellcolor{gray!5}18 & \cellcolor{gray!5}35 & \cellcolor{gray!5}3 & \cellcolor{gray!5}8 \cr
\cellcolor{gray!3}20. Uber & \cellcolor{gray!3}13 & \cellcolor{gray!3}61 & \cellcolor{blue!2}15 & \cellcolor{blue!2}38 & \cellcolor{blue!2}3 & \cellcolor{blue!2}5 & \cellcolor{green!2}17 & \cellcolor{green!2}30 & \cellcolor{green!2}5 & \cellcolor{green!2}9 & \cellcolor{purple!2}12 & \cellcolor{purple!2}32 & \cellcolor{purple!2}9 & \cellcolor{purple!2}8 & \cellcolor{gray!2}14 & \cellcolor{gray!2}33 & \cellcolor{gray!2}2 & \cellcolor{gray!2}11 \cr
\cellcolor{gray!8}21. Shazam & \cellcolor{gray!8}9 & \cellcolor{gray!8}87 & \cellcolor{blue!5}24 & \cellcolor{blue!5}54 & \cellcolor{blue!5}5 & \cellcolor{blue!5}4 & \cellcolor{green!5}11 & \cellcolor{green!5}51 & \cellcolor{green!5}11 & \cellcolor{green!5}14 & \cellcolor{purple!5}17 & \cellcolor{purple!5}58 & \cellcolor{purple!5}7 & \cellcolor{purple!5}5 & \cellcolor{gray!5}20 & \cellcolor{gray!5}47 & \cellcolor{gray!5}7 & \cellcolor{gray!5}13 \cr
\cellcolor{gray!3}22. Grab & \cellcolor{gray!3}9 & \cellcolor{gray!3}74 & \cellcolor{blue!2}17 & \cellcolor{blue!2}50 & \cellcolor{blue!2}4 & \cellcolor{blue!2}3 & \cellcolor{green!2}11 & \cellcolor{green!2}50 & \cellcolor{green!2}4 & \cellcolor{green!2}9 & \cellcolor{purple!2}9 & \cellcolor{purple!2}54 & \cellcolor{purple!2}4 & \cellcolor{purple!2}7 & \cellcolor{gray!2}18 & \cellcolor{gray!2}47 & \cellcolor{gray!2}4 & \cellcolor{gray!2}5 \cr
\cellcolor{gray!8}23. Spotify & \cellcolor{gray!8}10 & \cellcolor{gray!8}68 & \cellcolor{blue!5}25 & \cellcolor{blue!5}33 & \cellcolor{blue!5}6 & \cellcolor{blue!5}4 & \cellcolor{green!5}9 & \cellcolor{green!5}45 & \cellcolor{green!5}6 & \cellcolor{green!5}8 & \cellcolor{purple!5}11 & \cellcolor{purple!5}47 & \cellcolor{purple!5}7 & \cellcolor{purple!5}3 & \cellcolor{gray!5}24 & \cellcolor{gray!5}32 & \cellcolor{gray!5}2 & \cellcolor{gray!5}10 \cr
\cellcolor{gray!3}24. Canva & \cellcolor{gray!3}11 & \cellcolor{gray!3}73 & \cellcolor{blue!2}21 & \cellcolor{blue!2}41 & \cellcolor{blue!2}5 & \cellcolor{blue!2}6 & \cellcolor{green!2}9 & \cellcolor{green!2}54 & \cellcolor{green!2}3 & \cellcolor{green!2}7 & \cellcolor{purple!2}13 & \cellcolor{purple!2}49 & \cellcolor{purple!2}8 & \cellcolor{purple!2}3 & \cellcolor{gray!2}20 & \cellcolor{gray!2}42 & \cellcolor{gray!2}7 & \cellcolor{gray!2}4 \cr

\cellcolor{gray!8}25. Yahho Sports & \cellcolor{gray!8}8 & \cellcolor{gray!8}92 & \cellcolor{blue!5}14 & \cellcolor{blue!5}7 & \cellcolor{blue!5}4 & \cellcolor{blue!5}6 & \cellcolor{green!5}11 & \cellcolor{green!5}12 & \cellcolor{green!5}4 & \cellcolor{green!5}4 & \cellcolor{purple!5}6 & \cellcolor{purple!5}9 & \cellcolor{purple!5}3 & \cellcolor{purple!5}3 &\cellcolor{gray!5}10 & \cellcolor{gray!5}3 & \cellcolor{gray!5}4 & \cellcolor{gray!5}11 \cr
\cellcolor{gray!3}26. ESPN & \cellcolor{gray!3}10 & \cellcolor{gray!3}114 & \cellcolor{blue!2}45 & \cellcolor{blue!2}52 & \cellcolor{blue!2}8 & \cellcolor{blue!2}10 & \cellcolor{green!2}33 & \cellcolor{green!2}51 & \cellcolor{green!2}9 & \cellcolor{green!2}25 & \cellcolor{purple!2}31 & \cellcolor{purple!2}41 & \cellcolor{purple!2}19 & \cellcolor{purple!2}25 & \cellcolor{gray!2}36 & \cellcolor{gray!2}50 & \cellcolor{gray!2}10 & \cellcolor{gray!2}23 \cr
\cellcolor{gray!8}27. Wechat & \cellcolor{gray!8}8 & \cellcolor{gray!8}101 & \cellcolor{blue!5}24 & \cellcolor{blue!5}61 & \cellcolor{blue!5}8 & \cellcolor{blue!5}8 & \cellcolor{green!5}25 & \cellcolor{green!5}63 & \cellcolor{green!5}6 & \cellcolor{green!5}11 & \cellcolor{purple!5}28 & \cellcolor{purple!5}51 & \cellcolor{purple!5}18 & \cellcolor{purple!5}4 & \cellcolor{gray!5}11 & \cellcolor{gray!5}58 & \cellcolor{gray!5}11 & \cellcolor{gray!5}26 \cr
\cellcolor{gray!3}28. NBC Sports & \cellcolor{gray!3}4 & \cellcolor{gray!3}42 & \cellcolor{blue!2}7 & \cellcolor{blue!2}25 & \cellcolor{blue!2}6 & \cellcolor{blue!2}4 & \cellcolor{green!2}11 & \cellcolor{green!2}27 & \cellcolor{green!2}0 & \cellcolor{green!2}13 & \cellcolor{purple!2}22 & \cellcolor{purple!2}9 & \cellcolor{purple!2}2 & \cellcolor{purple!2}2 & \cellcolor{gray!2}9 & \cellcolor{gray!2}24 & \cellcolor{gray!2}7 & \cellcolor{gray!2}8 \cr
\cellcolor{gray!8}29. YouTube & \cellcolor{gray!8}11 & \cellcolor{gray!8}172 & \cellcolor{blue!5}38 & \cellcolor{blue!5}130 & \cellcolor{blue!5}3 & \cellcolor{blue!5}1 & \cellcolor{green!5}30 & \cellcolor{green!5}128 & \cellcolor{green!5}5 & \cellcolor{green!5}15 & \cellcolor{purple!5}32 & \cellcolor{purple!5}107 & \cellcolor{purple!5}26 & \cellcolor{purple!5}7 & \cellcolor{gray!5}34 & \cellcolor{gray!5}121 & \cellcolor{gray!5}12 & \cellcolor{gray!5}10 \cr
\cellcolor{gray!3}30. Facebook & \cellcolor{gray!3}14 & \cellcolor{gray!3}232 & \cellcolor{blue!2}42 & \cellcolor{blue!2}176 & \cellcolor{blue!2}9 & \cellcolor{blue!2}5 & \cellcolor{green!2}40 & \cellcolor{green!2}177 & \cellcolor{green!2}8 & \cellcolor{green!2}7 & \cellcolor{purple!2}37 & \cellcolor{purple!2}157 & \cellcolor{purple!2}28 & \cellcolor{purple!2}10 & \cellcolor{gray!2}42 & \cellcolor{gray!2}170 & \cellcolor{gray!2}15 & \cellcolor{gray!2}12 \cr
\cellcolor{gray!8}31. Twitter & \cellcolor{gray!8}11 & \cellcolor{gray!8}192 & \cellcolor{blue!5}16 & \cellcolor{blue!5}155 & \cellcolor{blue!5}13 & \cellcolor{blue!5}8 & \cellcolor{green!5}22 & \cellcolor{green!5}160 & \cellcolor{green!5}8 & \cellcolor{green!5}8 & \cellcolor{purple!5}18 & \cellcolor{purple!5}136 & \cellcolor{purple!5}32 & \cellcolor{purple!5}6 & \cellcolor{gray!5}21 & \cellcolor{gray!5}145 & \cellcolor{gray!5}23 & \cellcolor{gray!5}8 \cr
\cellcolor{gray!3}32. Telegram & \cellcolor{gray!3}9 & \cellcolor{gray!3}101 & \cellcolor{blue!2}11 & \cellcolor{blue!2}78 & \cellcolor{blue!2}6 & \cellcolor{blue!2}6 & \cellcolor{green!2}12 & \cellcolor{green!2}74 & \cellcolor{green!2}10 & \cellcolor{green!2}11 & \cellcolor{purple!2}14 & \cellcolor{purple!2}61 & \cellcolor{purple!2}23 & \cellcolor{purple!2}3 & \cellcolor{gray!2}12 & \cellcolor{gray!2}70 & \cellcolor{gray!2}14 & \cellcolor{gray!2}11 \cr
\cellcolor{gray!8}33. Whatsapp & \cellcolor{gray!8}8 & \cellcolor{gray!8}97 & \cellcolor{blue!5}17 & \cellcolor{blue!5}71 & \cellcolor{blue!5}5 & \cellcolor{blue!5}4 & \cellcolor{green!5}19 & \cellcolor{green!5}72 & \cellcolor{green!5}4 & \cellcolor{green!5}3 & \cellcolor{purple!5}14 & \cellcolor{purple!5}62 & \cellcolor{purple!5}14 & \cellcolor{purple!5}6 & \cellcolor{gray!5}18 & \cellcolor{gray!5}70 & \cellcolor{gray!5}6 & \cellcolor{gray!5}9 \cr
\cellcolor{gray!3}34. Google Chat & \cellcolor{gray!3}7 & \cellcolor{gray!3}66 & \cellcolor{blue!2}22 & \cellcolor{blue!2}39 & \cellcolor{blue!2}3 & \cellcolor{blue!2}3 & \cellcolor{green!2}16 & \cellcolor{green!2}39 & \cellcolor{green!2}3 & \cellcolor{green!2}9 & \cellcolor{purple!2}7 & \cellcolor{purple!2}30 & \cellcolor{purple!2}12 & \cellcolor{purple!2}17 & \cellcolor{gray!2}12 & \cellcolor{gray!2}35 & \cellcolor{gray!2}7 & \cellcolor{gray!2}17 \cr
\cellcolor{gray!8}35. Skype & \cellcolor{gray!8}6 & \cellcolor{gray!8}55 & \cellcolor{blue!5}14 & \cellcolor{blue!5}26 & \cellcolor{blue!5}7 & \cellcolor{blue!5}8 & \cellcolor{green!5}10 & \cellcolor{green!5}29 & \cellcolor{green!5}4 & \cellcolor{green!5}15 & \cellcolor{purple!5}10 & \cellcolor{purple!5}21 & \cellcolor{purple!5}12 & \cellcolor{purple!5}13 & \cellcolor{gray!5}12 & \cellcolor{gray!5}27 & \cellcolor{gray!5}6 & \cellcolor{gray!5}19 \cr
\cellcolor{gray!3}36. Google LLC & \cellcolor{gray!3}7 & \cellcolor{gray!3}184 & \cellcolor{blue!2}80 & \cellcolor{blue!2}91 & \cellcolor{blue!2}6 & \cellcolor{blue!2}8 & \cellcolor{green!2}69 & \cellcolor{green!2}93 & \cellcolor{green!2}4 & \cellcolor{green!2}35 & \cellcolor{purple!2}50 & \cellcolor{purple!2}84 & \cellcolor{purple!2}13 & \cellcolor{purple!2}41 & \cellcolor{gray!2}75 & \cellcolor{gray!2}90 & \cellcolor{gray!2}7 & \cellcolor{gray!2}24 \cr
\cellcolor{gray!8}37. PayPal & \cellcolor{gray!8}5 & \cellcolor{gray!8}51 & \cellcolor{blue!5}14 & \cellcolor{blue!5}27 & \cellcolor{blue!5}5 & \cellcolor{blue!5}5 & \cellcolor{green!5}12 & \cellcolor{green!5}30 & \cellcolor{green!5}2 & \cellcolor{green!5}10 & \cellcolor{purple!5}13 & \cellcolor{purple!5}24 & \cellcolor{purple!5}8 & \cellcolor{purple!5}6 & \cellcolor{gray!5}11 & \cellcolor{gray!5}29 & \cellcolor{gray!5}3 & \cellcolor{gray!5}16 \cr
\cellcolor{gray!3}38. Messenger & \cellcolor{gray!3}9 & \cellcolor{gray!3}144 & \cellcolor{blue!2}44 & \cellcolor{blue!2}83 & \cellcolor{blue!2}7 & \cellcolor{blue!2}10 & \cellcolor{green!2}32 & \cellcolor{green!2}87 & \cellcolor{green!2}3 & \cellcolor{green!2}26 & \cellcolor{purple!2}13 & \cellcolor{purple!2}73 & \cellcolor{purple!2}17 & \cellcolor{purple!2}41 & \cellcolor{gray!2}40 & \cellcolor{gray!2}82 & \cellcolor{gray!2}8 & \cellcolor{gray!2}21 \cr
\cellcolor{gray!8}39. LINE & \cellcolor{gray!8}8 & \cellcolor{gray!8}94 & \cellcolor{blue!5}18 & \cellcolor{blue!5}62 & \cellcolor{blue!5}8 & \cellcolor{blue!5}6 & \cellcolor{green!5}20 & \cellcolor{green!5}64 & \cellcolor{green!5}6 & \cellcolor{green!5}4 & \cellcolor{purple!5}4 & \cellcolor{purple!5}51 & \cellcolor{purple!5}19 & \cellcolor{purple!5}20 & \cellcolor{gray!5}22 & \cellcolor{gray!5}60 & \cellcolor{gray!5}10 & \cellcolor{gray!5}7 \cr
\cellcolor{gray!3}40. Gmail & \cellcolor{gray!3}4 & \cellcolor{gray!3}22 & \cellcolor{blue!2}8 & \cellcolor{blue!2}12 & \cellcolor{blue!2}0 & \cellcolor{blue!2}2 & \cellcolor{green!2}5 & \cellcolor{green!2}12 & \cellcolor{green!2}0 & \cellcolor{green!2}5 & \cellcolor{purple!2}6 & \cellcolor{purple!2}7 & \cellcolor{purple!2}5 & \cellcolor{purple!2}4 & \cellcolor{gray!2}6 & \cellcolor{gray!2}11 & \cellcolor{gray!2}1 & \cellcolor{gray!2}12 \cr

\cellcolor{gray!8}41. Maverick & \cellcolor{gray!8}6 & \cellcolor{gray!8}60 & \cellcolor{blue!5}19 & \cellcolor{blue!5}34 & \cellcolor{blue!5}2 & \cellcolor{blue!5}5 & \cellcolor{green!5}13 & \cellcolor{green!5}36 & \cellcolor{green!5}3 & \cellcolor{green!5}8 & \cellcolor{purple!5}10 & \cellcolor{purple!5}35 & \cellcolor{purple!5}11 & \cellcolor{purple!5}4 & \cellcolor{gray!5}19 & \cellcolor{gray!5}31 & \cellcolor{gray!5}3 & \cellcolor{gray!5}7 \cr
\cellcolor{gray!3}42. TimTree & \cellcolor{gray!3}8 & \cellcolor{gray!3}47 & \cellcolor{blue!2}8 & \cellcolor{blue!2}34 & \cellcolor{blue!2}1 & \cellcolor{blue!2}4 & \cellcolor{green!2}7 & \cellcolor{green!2}29 & \cellcolor{green!2}2 & \cellcolor{green!2}9 & \cellcolor{purple!2}7 & \cellcolor{purple!2}19 & \cellcolor{purple!2}9 & \cellcolor{purple!2}12 & \cellcolor{gray!2}8 & \cellcolor{gray!2}27 & \cellcolor{gray!2}2 & \cellcolor{gray!2}10 \cr
\cellcolor{gray!8}43. Vinted & \cellcolor{gray!8}8 & \cellcolor{gray!8}57 & \cellcolor{blue!5}26 & \cellcolor{blue!5}23 & \cellcolor{blue!5}3 & \cellcolor{blue!5}5 & \cellcolor{green!5}22 & \cellcolor{green!5}18 & \cellcolor{green!5}4 & \cellcolor{green!5}13 & \cellcolor{purple!5}18 & \cellcolor{purple!5}21 & \cellcolor{purple!5}7 & \cellcolor{purple!5}11 & \cellcolor{gray!5}25 & \cellcolor{gray!5}22 & \cellcolor{gray!5}1 & \cellcolor{gray!5}9 \cr
\cellcolor{gray!3}44. Allset & \cellcolor{gray!3}7 & \cellcolor{gray!3}94 & \cellcolor{blue!2}17 & \cellcolor{blue!2}67 & \cellcolor{blue!2}3 & \cellcolor{blue!2}7 & \cellcolor{green!2}8 & \cellcolor{green!2}62 & \cellcolor{green!2}7 & \cellcolor{green!2}17 & \cellcolor{purple!2}11 & \cellcolor{purple!2}67 & \cellcolor{purple!2}9 & \cellcolor{purple!2}7 & \cellcolor{gray!2}19 & \cellcolor{gray!2}66 & \cellcolor{gray!2}2 & \cellcolor{gray!2}7 \cr
\cellcolor{gray!8}45. Headway & \cellcolor{gray!8}9 & \cellcolor{gray!8}98 & \cellcolor{blue!5}25 & \cellcolor{blue!5}60 & \cellcolor{blue!5}4 & \cellcolor{blue!5}9 & \cellcolor{green!5}17 & \cellcolor{green!5}60 & \cellcolor{green!5}10 & \cellcolor{green!5}21 & \cellcolor{purple!5}19 & \cellcolor{purple!5}73 & \cellcolor{purple!5}6 & \cellcolor{purple!5}10 & \cellcolor{gray!5}21 & \cellcolor{gray!5}69 & \cellcolor{gray!5}3 & \cellcolor{gray!5}15 \cr
\cellcolor{gray!3}46. OpenSea & \cellcolor{gray!3}11 & \cellcolor{gray!3}78 & \cellcolor{blue!2}23 & \cellcolor{blue!2}37 & \cellcolor{blue!2}3 & \cellcolor{blue!2}4 & \cellcolor{green!2}23 & \cellcolor{green!2}43 & \cellcolor{green!2}5 & \cellcolor{green!2}7 & \cellcolor{purple!2}14 & \cellcolor{purple!2}52 & \cellcolor{purple!2}5 & \cellcolor{purple!2}7 & \cellcolor{gray!2}21 & \cellcolor{gray!2}44 & \cellcolor{gray!2}1 & \cellcolor{gray!2}12 \cr
\cellcolor{gray!8}47. Klarna & \cellcolor{gray!8}9 & \cellcolor{gray!8}79 & \cellcolor{blue!5}20 & \cellcolor{blue!5}52 & \cellcolor{blue!5}2 & \cellcolor{blue!5}5 & \cellcolor{green!5}8 & \cellcolor{green!5}45 & \cellcolor{green!5}9 & \cellcolor{green!5}17 & \cellcolor{purple!5}10 & \cellcolor{purple!5}51 & \cellcolor{purple!5}12 & \cellcolor{purple!5}6 & \cellcolor{gray!5}19 & \cellcolor{gray!5}42 & \cellcolor{gray!5}8 & \cellcolor{gray!5}10 \cr
\cellcolor{gray!3}48. SideChef & \cellcolor{gray!3}7 & \cellcolor{gray!3}59 & \cellcolor{blue!2}19 & \cellcolor{blue!2}33 & \cellcolor{blue!2}4 & \cellcolor{blue!2}3 & \cellcolor{green!2}15 & \cellcolor{green!2}31 & \cellcolor{green!2}5 & \cellcolor{green!2}8 & \cellcolor{purple!2}7 & \cellcolor{purple!2}40 & \cellcolor{purple!2}8 & \cellcolor{purple!2}4 & \cellcolor{gray!2}19 & \cellcolor{gray!2}33 & \cellcolor{gray!2}2 & \cellcolor{gray!2}5 \cr
\hline
\cellcolor{gray!8}Metric & \cellcolor{gray!8} & \cellcolor{gray!8} & \cellcolor{blue!5} & \cellcolor{blue!5}P  & \cellcolor{blue!5}R & \cellcolor{blue!5}F1 & \cellcolor{green!5} & \cellcolor{green!5}P & \cellcolor{green!5}R & \cellcolor{green!5}F1 & \cellcolor{purple!5} & \cellcolor{purple!5}P & \cellcolor{purple!5}R & \cellcolor{purple!5}F1 & \cellcolor{gray!5} & \cellcolor{gray!5}P & \cellcolor{gray!5}R & \cellcolor{gray!5}F1 \cr
\cellcolor{gray!3}Average value \textcolor{gray}{(\%)} & \cellcolor{gray!3} & \cellcolor{gray!3} & \cellcolor{blue!2} & \cellcolor{blue!2}\textbf{83.5} & \cellcolor{blue!2}\textbf{78.9} & \cellcolor{blue!2}\textbf{81.2} & \cellcolor{green!2} & \cellcolor{green!2}75.3 & \cellcolor{green!2}63.4 & \cellcolor{green!2}68.8 & \cellcolor{purple!2} & \cellcolor{purple!2}59.6 & \cellcolor{purple!2}63.9 & \cellcolor{purple!2}61.7 & \cellcolor{gray!2} & \cellcolor{gray!2}78.3 & \cellcolor{gray!2}63.4 & \cellcolor{gray!2}70.1 \cr
\hline
\end{tabular}
\end{center}
\end{table}

\subsection{RQ1: Evaluation of the effectiveness}\label{sub: effectiveness}
To answer this research question, we conduct two experiments.
The first one involves evaluating the effectiveness of ALVIN using real-world apps and comparing it with other methods. 
The second experiment is an ablation study, aiming at investigating the impact of removing invisible views from the GUIs on the effectiveness of ALVIN.

\subsubsection{Effectiveness of ALVIN}
In this experiment, the objective is to assess how effective ALVIN is, and compare it with other baseline tools to reveal its advantages.
\\
\textbf{\underline{Data preparation:}} We start by preparing datasets for training the model and evaluating the effectiveness of ALVIN.
The list of recommendations in Google Play is high downloads, highly rated apps, which are also easy to be accepted by these users~\citep{Alshayban2020AccessibilityII}.
We then sequentially crawl the \emph{.apk} files of recommended apps in six domains from their index page on April 7, 2023, and install these apps in the Google Nexus 6 on the emulator with Android 11.0 OS.
A total of 500 apps are collected for this research question, involving \emph{Education} (96 apps), \emph{Sports} (64 apps), \emph{Book} (100 apps), \emph{Food} (110 apps), \emph{Map} (72 apps), and \emph{Medical} (58 apps). 

Then, we extract GUIs of each app using the AppCrawler~\citep{appCrawler}, and parse their layout-related files (\emph{*.uix}) by the uiautomator~\citep{UIautomator}.
AppCrawler could simulate the operation of apps, and take screenshots in GUIs.
However, this tool cannot traverse GUIs that require users to log in (e.g., the account pages), existing pop-up windows, and other unexpected situations.
Therefore, we only collected a subset of GUIs in the apps. 
Afterward, we manually remove those GUIs that only contain one or two components, and duplicate items. 
In total, there are 2,646 available GUIs left behind. 
We then regard the visible \emph{$<$imageviews$>$}, \emph{$<$buttonviews$>$}, \emph{$<$textviews$>$}, \emph{$<$listviews$>$}, and \emph{$<$searchviews$>$} as the components.
Besides, there are components that might be overlayed by multiple views, and we would treat them as one component.
After that, we totally collect 88,570 components.

From the special education school, we recruit 6 low vision users to assist us in annotating the accessibility issues in GUIs.
Instead of marking these components separately in parallel and checking the consistency of labels, these users jointly analyze the components and conduct the discussion.
The first and fifth authors of this paper would manually record the components and their accessibility issues in real-time, a total of 400 GUIs were annotated in this way. 
However, it is an energy-consuming task for them to annotate all GUIs.
Therefore, we further recruit 10 highly nearsighted volunteers from our college.
We ask them to take off their glasses, adjust the distance between the screen and their eyes, and exercise each of them to re-annotate 10 GUIs that low vision users have finished before, until the results are almost identical.
These volunteers could then annotate the remaining GUIs in our datasets as they did during the exercise.
Notably, the GUIs annotated by these volunteers are independent of each other, averaging about 225 GUIs per person, and each volunteer has two weeks to complete this task, which would not affect the results due to excessive fatigue.
The datasets we annotate, although not absolutely represent the opinions of all low vision users, we argue that they are in line with their practical experience.
Quantitatively, 91\% of the final discussion results correspond to the most frequently occurring initial annotations, while the remaining 9\% are decisions revised through discussion. 
This demonstrates a high level of initial agreement and the effectiveness of the discussion process in resolving discrepancies.

In practice, it is a huge task for us to check all the 500 apps we collected, so we intend to randomly select a subset of apps to construct the ground truth and evaluate our tool. 
Then, following the works~\citep{Bajammal2021SemanticWA, Li2022PushButtonSO, Xie2022PsychologicallyinspiredUI}, in which all selected subsets from their total dataset to evaluate the effectiveness of their method, we then select 32 apps based on the downloads.
Further, to clarify the generalization of our method, we also randomly selected 16 apps from Google Play to participate in the experiment, so we test 48 apps in total and divide them into two groups as follows.
One group contains 24 apps that are above the median downloads (25 million in our datasets) in the first half of Table~\ref{tab: effectiveness}, and the other 24 apps are below this value shown in the other half of Table~\ref{tab: effectiveness}.
We then regard such selected apps as our ground truths to evaluate the effectiveness of ALVIN.
Among the other 468 apps, 80\% of them belong to the training set of the multi-classification model as we mentioned in Section~\ref{sub: GCN}, while the other 20\% are for the validation set. 
Notably, in this experiment, we examine all views and structural information of each GUI layout file to locate inaccessible components.
\\
\textbf{\underline{Baselines:}} In order to have a more thorough evaluation, we include 3 baselines in our experiments.
Accessibility Scanner (AS)~\citep{AccessibilityScanner}, released by Google Research, can scan the GUIs and provide suggestions to improve the accessibility of apps, based on content labels, touch target size, clickable items, and text and image contrast.
In detail, the AS can detect small sizes of components, narrow intervals, low color contrast, lack of content-description, and focusability of components in a GUI using pre-set decision rules (i.e., the size of the component must not be less than $24px \times 24px$) for a static GUI screenshot or a continuous GUI operation record. 
Of these, the first three are relevant to the low vision users for whom our work is intended, while the last two are usually intended for blind users.
Mobile Automated Test Framework (MATE)~\citep{Eler2018AutomatedAT}, proposed by Eler et al., supports checking the accessibility issues in GUIs by formulating rules, and providing related reports.
Regarding this baseline, it used constructed rules to detect and identify three accessibility issues, small component size, narrow interval, and the missing of content-description, by analyzing the GUI layout files. 
Its only issues relevant to low vision users are the first two.
Xbot~\citep{Chen2021AccessibleON}, designed by Chen et al., can capture various accessibility issues in GUIs by optimizing existing rules, and it can collect more issues.
This tool was developed based on Google Accessibility Test Framework~\footnote{\url{https://github.com/google/Accessibility-Test-Framework-for-Android}} and was further extended and optimized by extracting the required activity intent parameters for launching each activity to make the tool more focused on GUI accessibility issues. 
In total, it involved the detection of the following types of accessibility issues: component size, component interval, text and image contrast, component label, and unapparent link.
The rationale for selecting these three baselines is that they perform relatively well among the existing known tools~\citep{Zhang2018RobustAO, Oliveira2016AccessibilityOT, Zhang2021ScreenRC, Hao2014PUMAPU, MobileAccessibilityChecker} after our exploration, and either from well-known companies or from recent research, which are more convincing than others.
\\
\textbf{\underline{Methodology:}} To evaluate the effectiveness, we apply ALVIN to the GUIs in our ground truth, as well as re-running the above baseline tools on the same GUIs.
In this process, we would debug, until all GUIs work in the aforementioned tools well.
Since the baseline tools would capture the blind-related accessibility issues (e.g., lacking content-description), we filter these out to ensure the fairness of the experiments.

Subsequently, we only focus on whether the inaccessible components could be effectively identified without considering what types they have been classified. Thus, we categorize each result into one of the following:
\textbf{True positive (TP):} It refers to the components that are classified by ALVIN for the inaccessible components and consistent with their labels we manually annotate. 
\textbf{True negative (TN):} This represents the components that are classified with general labels, as well as the general components that we manually annotate. 
\textbf{False positive (FP):} This case is the components classified as inaccessible, but low vision users deem they are general. 
\textbf{False negative (FN):} This item is the components that are classified as general components, but they suffer from accessibility issues in our ground truths.
We further adopt the number of TP, TN, FP, and FN to calculate the precision, recall, and F1-score, respectively.
\\
\textbf{\underline{Results and discussion:}} Table~\ref{tab: effectiveness} shows the results of evaluating the effectiveness of ALVIN.
The columns present the name of apps, the number of GUIs intercepted from these apps, the number of their components (Coms), and the number of TP, TN, FP, and FN of ALVIN and baseline tools.
The last row shows the average value of precision, recall, and F1-score.

The key outcome of this evaluation is P (=83.5\%), R (=78.9\%), and F1 (=81.2\%) of ALVIN all exceed other baseline tools, seeing the bold contents in Table~\ref{tab: effectiveness}.
These results indicate a rather effective performance of ALVIN.
The rationale for ALVIN outperforming the baseline tools is that such a GCN-based method could identify the relations between components well, and effectively prevent inaccessible components from being omitted compared with the rule-based methods.
For instance, when multiple components belong to the same container, ALVIN could retain the relations among them, so that once there is an inaccessible component, the possibility that others will be annotated as such increases accordingly.
On both sides of the median in app downloads, there are varying amounts of accessibility issues among apps.
For the first half of apps (downloads $\textless$ 25 million) in Table~\ref{tab: effectiveness}, an average of 48.8\% of components have accessibility issues.
In contrast, the other half of apps (downloads $\textgreater$ 25 million) in Table~\ref{tab: effectiveness} possess 17.7\% of inaccessible components on average.
This observation is expected since highly downloaded apps are more likely to have more resources to create better products.
In practice, these baseline tools might capture many invisible components, and ignore or annotate extra components, like the blank views in the GUIs.
The rationale for such problematic results is that the baseline tools are all rule-based and are enforced to analyze all information in the layout files.
These troubles will hinder developers from effectively locating inaccessible components and affect the subsequent fixed process.
Meanwhile, this is also the main reason for the baseline tools perform low F1-score.

We then investigate the reasons behind the false positives and negatives.
Typically, ALVIN encounters failures in scenarios when one component without accessibility issues is associated with multiple other components that have accessibility issues, or vice versa. 
Under the influence of such relationships, a component that should not have any issues may be incorrectly assigned the same accessibility issues as its associated components if those associated components have issues.
Essentially, this situation can be attributed to developers' non-standard design of GUI components. 
Developers may intentionally reduce the size or color contrast of other components to entice users to click on a specific component, resulting in visual discrepancies among components that should ideally share a uniform design style. 
This discrepancy makes it challenging for our method to accurately identify the accessibility issues.
The selected apps in our ground truths have at least 8 million downloads, but the lack of such apps with quite low downloads illustrates whether ALVIN could be applied to various apps well.
We argue that these apps are normally used by low vision users, so the examination of these apps could be more in line with the practical experience of these users, and also reveal the effectiveness of ALVIN on a certain level.

\begin{figure*}
\centering
\includegraphics[width=14.5cm]{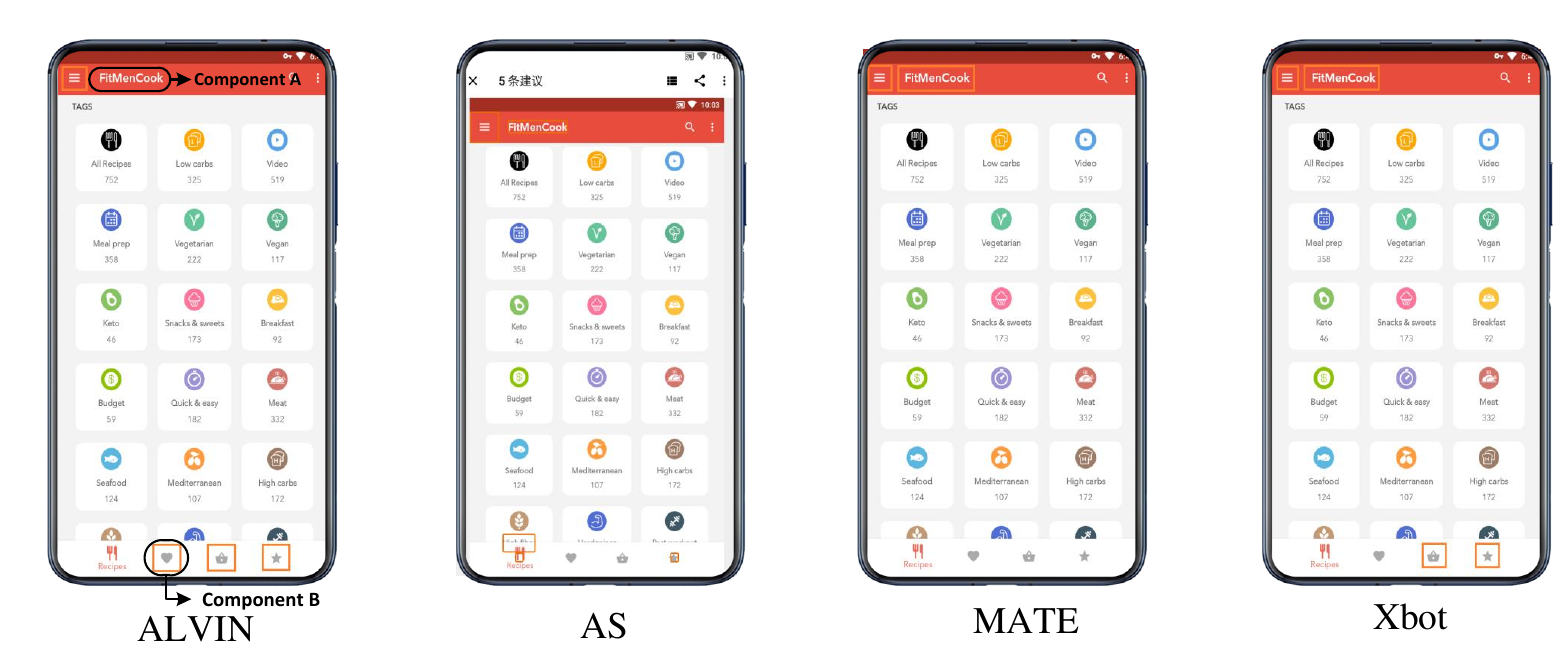}
\caption{The visual results of ALVIN and other baseline methods.}
\label{fig: vresults}
\end{figure*}
Figure~\ref{fig: vresults} presents the visual performance of different methods for capturing the GUI accessibility issues.
Overall, ALVIN could check the accessibility issues in the GUIs more accurately than others, and is more valuable for developers to fix these issues.
We further explain ALVIN's advantages in detail with two examples of component $A$ and $B$ signed in Figure~\ref{fig: vresults} (ALVIN).
Component $A$ is a \emph{textview} with a size of $102px \times 15px$, which does not conform to the width should be more than $24px$ stipulated by the W3C~\citep{LowVision}.
Thus, the Accessibility Scanner, MATE, and Xbot all annotate $A$ as an inaccessible component, but ALVIN would not.
This is because ALVIN is developed following the actual experiences of low vision users.
In detail, these users can see $A$ relatively clearly and do not annotate this component as having accessibility issues, so ALVIN will not regard $A$ as an inaccessible component.
For the Component $B$ signed in Figure~\ref{fig: vresults} (ALVIN), it is located in the bottom navigation view of this GUI, and suffers from low color contrast like other components in this navigation.
However, these baseline tools only capture the accessibility issues with other components in this navigation, ignoring component $B$.
This is mainly because these rule-based methods might ignore the clickable \emph{$<$imageviews$>$}, like the component $B$.
While ALVIN could capture such components effectively, caused by these components are connected in the GUI-graph and affect each other, so if one component has accessibility issues, then there would be a higher ratio of the other components having the same issues.

More qualitatively, compared to baseline tools, ALVIN performs better in accurately detecting unclear alert information and identifying components with low color contrast issues. 
Additionally, ALVIN demonstrates better fault tolerance for components with small standard deviations. 
However, when dealing with highly customized or dynamic GUI components, the baseline tool may outperform ALVIN. 
For example, in a map app (named `Maverick'), the homepage has a dashboard component responsible for navigation guidance. 
This component is a highly customized element within this app domain, and unlike traditional images or text, it incorporates a unique combination of graphics and animations to convey directional information. 
Such a unique dynamic combination of components makes it difficult for ALVIN to effectively capture their attribute information, leading to incorrect detection. 
However, the Accessibility Scanner, which can take dynamic GUI operation records as input, is able to capture changes in the attributes of such components. 
This actually presents an opportunity to further enhance the generalizability of our method. 
By integrating other modalities such as animations or GIFs into the GUI-graphs, the comprehensiveness of our ALVIN might be significantly improved.

\subsubsection{Ablation study}\label{sub: ablation}
In this study, we design three ablation experiments: the first one investigates the impact of the attribute matrix on the model's performance when used as input, the second experiment examines the effect of the number of convolutional layers on the model's performance, and the third one explores the influence of fully connected layers.
Notably, to conduct these ablation experiments, we reorganize the dataset required for model training, validation, and testing. 
The dataset still consists of apps and GUIs that we have constructed in the previous experiment. 
\\
\textbf{\underline{Feature attribute matrix:}} In this experiment, we explore the influence of the feature attribute matrix we designed on the model's performance by computing the accuracy. 
Throughout this process, we refrain from directly removing the attribute matrix for experimentation, as it is evident that the model would hardly annotate any accessibility issues in the absence of this matrix. 
Therefore, we choose to remove accessibility attributes and inherent attributes from the attribute matrix one by one to explore the influence of these two types of attributes on the model's performance.
Notably, in both cases, we keep the label of the node, as this is the basic condition for the model to conduct multi-classification.
\\
\textbf{\underline{Results and discussion:}}The model's accuracy drops to 54.2\% upon the removal of accessibility attributes, while it stands at 70.3\% after removing inherent attributes. 
These results indicate that both types of attributes significantly impact the model's performance, but the accessibility attributes are more crucial. 
On one hand, the accessibility attributes are more numerous, providing a more comprehensive feature set for the nodes. 
On the other hand, the accessibility attributes include critical information such as the color (``intuitive'') of the GUI components and the alternative text (``text\_alternative''), which are essential for detecting accessibility issues of low color contrast and unclear alert information.
Further, the impact of inherent attributes on performance lies in their inclusion of the ``bounds'' attribute, which is essential for calculating component sizes and the intervals between components.
\begin{figure}
\centering
\includegraphics[width=8cm]{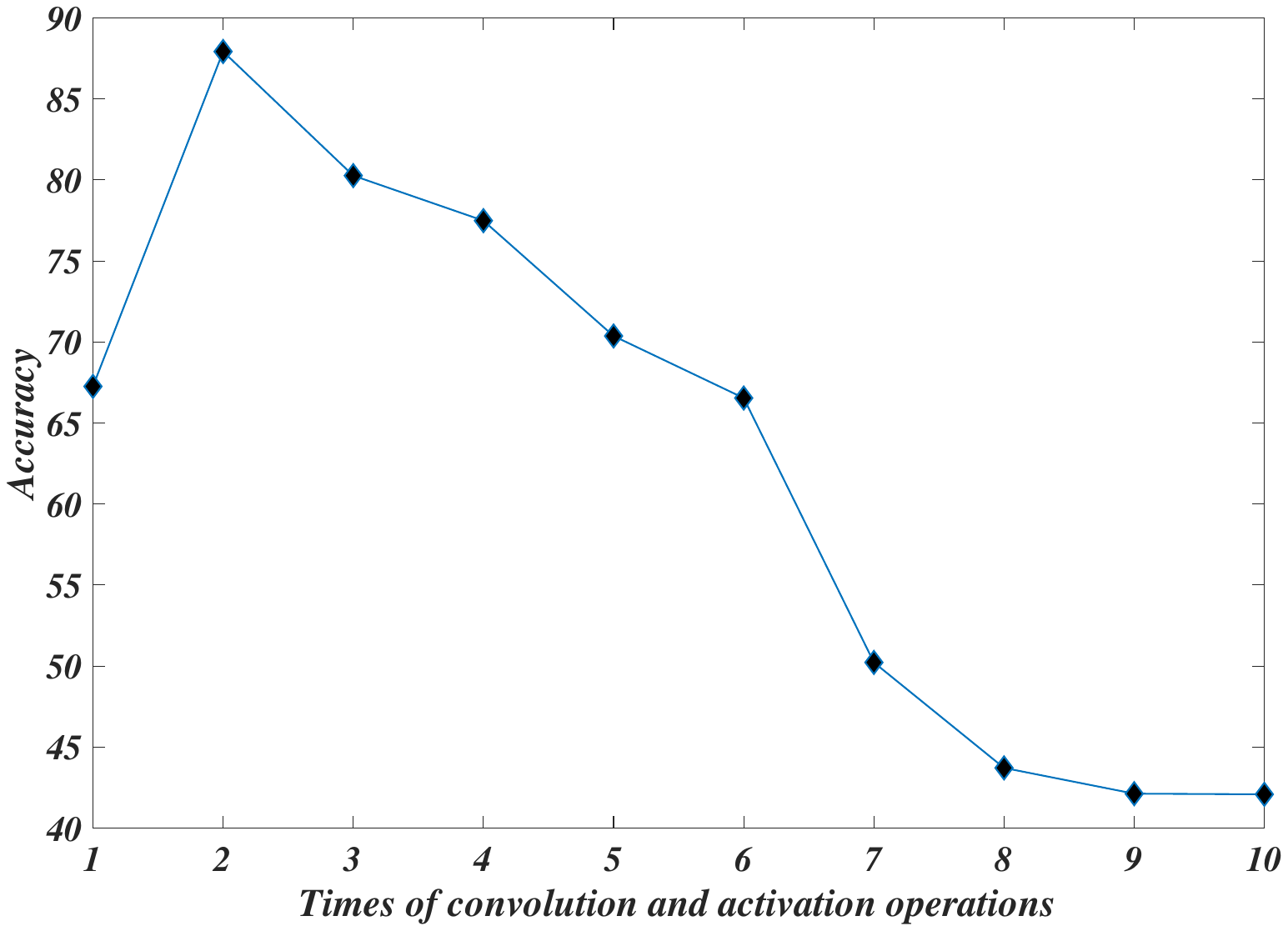}
\caption{The accuracy corresponds to the different number of convolutional layers.}
\label{fig: times}
\end{figure}
\\
\textbf{\underline{The number of convolutional layers:}} The purpose of this experiment is to discuss how many convolutional layers can make our GCN model optimal.
We start with one operation and iteratively increase the number, computing the accuracy of the detection results.
In this process, except for the changes in convolution times, other parameters are kept consistent to ensure the reliability of the results.
\\
\textbf{\underline{Results and discussion:}} Figure~\ref{fig: times} depicts the performance of the GCN model after ten iterations, which include one to ten times of convolution.
Where, the abscissa axis is the times of convolution, and the ordinate axis is the corresponding accuracy.
As we can see, in the beginning, the performance of the model continues to improve, but when the number of convolutions exceeds two, the performance of the model shows a significant downward trend.
Such a performance is in line with the ``degrade'' problem mentioned in~\citep{He2014ConvolutionalNN}~\citep{He2015DeepRL}.
This problem can be explained by the fact that adding extra layers to a shallow network, which has already reached its maximum capacity, may cause the network to degrade. 
If the weights of these layers are set to 1, the shallow network should perform the same as the deep network, as the extra layers have no effect on the network. 
However, training all the weights of the additional layers to be 1 is difficult, which may result in the degradation of the network when additional layers are added to the shallow network.
Based on these findings, we can conclude that the GCN model we built performs optimally with two convolutional layers, but adding more layers would be counterproductive.
\\
\textbf{\underline{Fully connected layer:}} In this experiment, we remove the fully connected layer of the model and instead opt to input the results of global max pooling into the softmax classifier for multi-class classification.
The purpose of this ablation experiment is to discuss the role of the fully connected layer in handling complex relationships within our model and to assess its impact on the model's performance.
\\
\textbf{\underline{Results and discussion:}} The experimental results indicate that, following the removal of the fully connected layer, the accuracy of ALVIN decreased from 87.93\% to 70.35\%. 
This can be attributed to two main reasons: firstly, the fully connected layer is capable of mapping input features into a high-dimensional space, extracting rich feature representations~\cite{AbuElHaija2019NGCNMG}. 
The removal of the fully connected layer may reduce the model's ability to extract features from the input data, leading to a decrease in feature representation capability; 
secondly, the fully connected layer possesses powerful parameterization capabilities, allowing it to learn complex nonlinear relationships among input features~\cite{kocsis2022unreasonable}. 
Removing the fully connected layer may weaken the model's ability to handle complex relationships, making it difficult to capture deep-level structures and associated information from the input data.

\begin{figure}
\centering
\includegraphics[width=8cm]{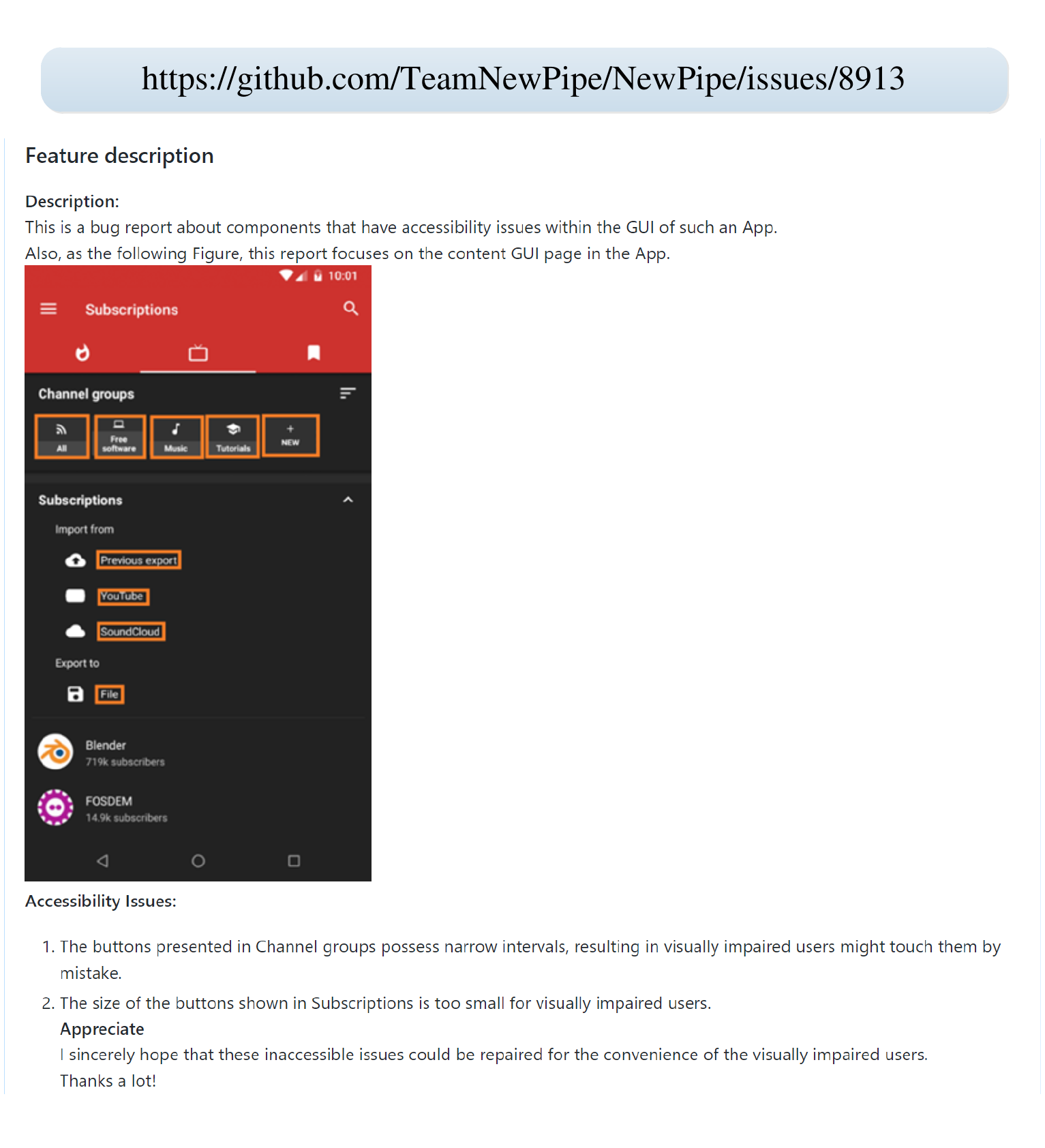}
\caption{An example of the submitted \emph{issue}.}
\label{fig: issue}
\end{figure}
\subsection{RQ2: Evaluation of the usefulness}\label{sub: usefulness}
In this research question, we focus on clarifying whether ALVIN is useful to guide developers in fixing their GUIs.
To this end, we first apply ALVIN to the open-source apps, and submit the issues report to the corresponding GitHub repositories.
Meanwhile, we record the comments from developers and whether they have fixed such issues.

F-Droid~\footnote{\url{https://f-droid.org/en/packages/}} provides open-source apps covering 17 different domains, but in order to fully consider the universality of ALVIN, we do not crawl apps from a single domain.
Also, owing to the ranking of apps in F-droid being constantly changing and irregular, we do not use the top apps directly.
Instead, we randomly crawl 20 open-source apps from the hot recommendation list, with their repositories hosted on GitHub.
Since these apps are well-maintained, the issues we submit could be handled promptly.
Also, such a random selection could avoid the potential influence of our subjective factors on the experimental results.

After applying ALVIN to the above apps, we create \emph{issues} from the generated report for the GUI page suffering the most accessibility issues, which is submitted to corresponding repositories.
In Figure~\ref{fig: issue}, we present an example of the \emph{issue} submitted to \emph{NewPipe}.
It consists of the following three parts: (1) the description, aiming to briefly introduce the purpose of this \emph{issue}, (2) the accessibility issues, we present the results and explain the identified accessibility issues in the GUI, (3) our appreciation to developers if they could notice and fix these issues in their apps.

We continually follow the submissions, given that developers may not understand what accessibility issues are and their effects on low vision users, we need to communicate with developers in time to explain their confusion.
In addition, developers may ask us if we could provide them with a complete and reusable code that can be used to fix these accessibility issues directly.
We explain that we would focus on automated fixing such accessibility issues in the future, and also assist them as much as possible. 
Notably, since developers might delete the \emph{issues} after fixing them, we take some screenshots of these \emph{issues} beforehand and save them in our open-sourced datasets.
\\
\textbf{\underline{Results and discussion:}} Table~\ref{tab: bugs} shows the detailed information and decisions of our submitted \emph{issues}, among such submitted 20 issues, 18 of them are successfully processed (8 fixed and 10 under fixing), while the other 2 issues are still pending since the reviewers have not maintained the repository for a long time.
In addition, we collect lots of feedback from developers through comments, and find that the majority of developers host a positive attitude towards improving the accessibility of apps.
This indicates that most developers sincerely expect more convenience visually impaired people could enjoy brought by their products equally.

\begin{table}\footnotesize
\renewcommand{\arraystretch}{1.3}
\caption{The decisions of our submitted \emph{issues}.\label{tab: bugs}}
\begin{center}
\begin{tabular}{lccc}
\hline
\rowcolor{gray!2}\textbf{apps}  & \textbf{Issue Id} & \textbf{Status (suc)} & \textbf{Issues} \cr
\hline
	\rowcolor{gray!8}1. AntennaPod & \#6196 & Fixed \textcolor{green}{$(\checkmark)$} & Interval, Color \cr
	\rowcolor{gray!3}2. Aard2 & \#150 & Fixing \textcolor{brown}{$(-)$} & Size \cr
	\rowcolor{gray!8}3. Baresip-studio & \#294 & Fixed \textcolor{green}{$(\checkmark)$} & Interval, Color \cr
	\rowcolor{gray!3}4. Cocoin & \#58 & Pending \textcolor{red}{$(\times)$} & Size, Interval \cr
	\rowcolor{gray!8}5. DroidRec & \#26 & Fixed \textcolor{green}{$(\checkmark)$} & Interval \cr
	\rowcolor{gray!3}6. Meshenger & \#88 & Fixed \textcolor{green}{$(\checkmark)$} & Size \cr 
	\rowcolor{gray!8}7. Koler & \#501 & Fixing \textcolor{brown}{$(-)$} & Size, Vague \cr
	\rowcolor{gray!3}8. ListMaker & \#9 & Fixed \textcolor{green}{$(\checkmark)$} & Color, Interval \cr
	\rowcolor{gray!8}9. Markdown & \#108 & Pending \textcolor{red}{$(\times)$} & Size, Color \cr
	\rowcolor{gray!3}10. Minimal-Todo & \#171 & Fixing \textcolor{brown}{$(-)$} & Size, Interval \cr
	\rowcolor{gray!8}11. Metronome & \#7 & Fixing \textcolor{brown}{$(-)$} & Color, Alert \cr
	\rowcolor{gray!3}12. Morse Trainer & \#224 & Fixing \textcolor{brown}{$(-)$} & Interval, Color \cr
	\rowcolor{gray!8}13. NewPipe & \#8913 & Fixed \textcolor{green}{$(\checkmark)$} & Interval, Vague \cr
	\rowcolor{gray!3}14. Shader Editor & \#107 & Fixing \textcolor{brown}{$(-)$} & Size, Interval \cr
	\rowcolor{gray!8}15. SuntimesWidget & \#656 & Fixed \textcolor{green}{$(\checkmark)$} & Size, Color \cr
	\rowcolor{gray!3}16. Syphon & \#250 & Fixing \textcolor{brown}{$(-)$} & Size, Color \cr
	\rowcolor{gray!8}17. Timber & \#473 & Fixing \textcolor{brown}{$(-)$} & Size, Color \cr
	\rowcolor{gray!3}18. TKCompanion & \#49 & Fixing \textcolor{brown}{$(-)$} & Interval, Color \cr
	\rowcolor{gray!8}19. Tuner & \#35 & Fixed \textcolor{green}{$(\checkmark)$} & Size \cr
	\rowcolor{gray!3}20. Tazapp-android & \#11 & Fixing \textcolor{brown}{$(-)$} & Size, Interval \cr
\hline
\end{tabular}
\end{center}
\end{table}
Among those processed \emph{issues}, the developers almost recognize the necessity of improving the accessibility of apps, and approve the results of ALVIN, a representative comment like \emph{\textcolor{c1}{``Thank you  much for figuring out these accessibility issues with the design of GUIs, we would be happy to repair them in subsequent releases.''}} 
These repositories have fixed their GUIs, indicating that ALVIN could work well.
Also, some developers mention that they do not understand what the GUI accessibility means, and how it affects visually impaired users.
One of them said, \emph{\textcolor{c1}{``Thanks for your comment. Please help me to understand better the motivation behind your issue. I will try to improve the situation, where it appears beneficial.''}} 
This suggests that developers might often ignore the accessibility of their apps, so more in-depth research efforts could well compensate for the lack of GUI accessibility.

Still, although some developers realize that their apps contain accessibility issues, they could not fix them due to technical and other realistic challenges.  
As one developer said that \emph{\textcolor{c1}{``Indeed, our app ignores accessibility-related standards in design, but it requires large-scale code refactoring.
We are grateful and agree with you, but it is hard for us.''}}
Therefore, if the research community can provide reliable automated tools for fixing accessibility issues, it would reduce the burden of code refactoring and modification for developers.

\vspace{-0.42em}
\subsection{RQ3: Evaluation of model selection}\label{sub: model}

In this research question, we discuss why we select the GCN model to implement ALVIN.
Thus, we elaborate on the performances of our adopted GCN model compared with Graph Attention Network (GAT)~\citep{Velickovic2017GraphAN}, Support Vector Machine (SVM)~\citep{Steinwart2008SupportVM}, Convolutional Neural Network (CNN)~\citep{LeCun2010ConvolutionalNA}, Random Forests (RF)~\citep{Breiman2001RandomF}, Multi-View Robust Graph Representation Learning (MGRL)~\citep{Ma2023MultiViewRG}, and Multi-Level Graph Relation Network (MuL-GRN)~\cite{Zhang2023MuLGRNMG}.
The reason for choosing these baseline models is that they all perform well on the task of node classification.

To this end, we input our constructed datasets (in Section~\ref{sub: effectiveness}) into these baselines, and further conduct multi-classification with the same labels as our GCN model.
However, there are two concerns with such baselines worth noting here.
One is that they require different forms of data, and the other is that the GAT model needs the self-attention mechanism~\citep{Vaswani2017AttentionIA}.
This mechanism is set between each node when shared weight changes linearly, and thus realizes the control of all edges in a shared manner.
As for the first concern, the dataset required for GAT is the same as GCN, so we adopt the same GUI-graphs to train this model.
This model is inputted with 1,929 GUI-graphs (80\% of the GUIs in 468 apps) as the training set, and the other 483 GUI-graphs (20\% of the GUIs in 468 apps) as the validation set.
Regarding the training of the GAT model, apart from the additional attention layers, its parameters and loss functions remain consistent with those used when employing GCN to ensure a fair comparison.
While the other baselines can only receive vectors to train the model and cannot process GUI-graphs.
Additionally, in the CNN, we set two convolutional layers and utilize the same activation and loss function as those employed in the GCN for training.

\begin{table}\scriptsize
\renewcommand{\arraystretch}{1.8}
  \centering
  \caption{Examples of predicting accessibility issues in different models}
  \begin{tabular}{ccccc}
    \hline
    \textbf{Items} & \textbf{Example 1} & \textbf{Example 2} & \textbf{Example 3} & \textbf{Example 4} \cr
    \hline
    \textbf{Examples}
    &     
    \begin{minipage}[b]{0.2\columnwidth}
		\centering
		\raisebox{-.5\height}{\includegraphics[width=\linewidth]{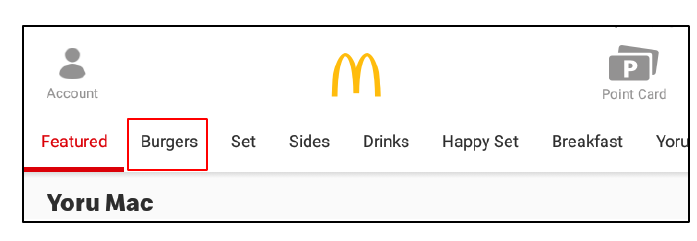}}
    \end{minipage}
    & 
    \begin{minipage}[b]{0.2\columnwidth}
		\centering
		\raisebox{-.5\height}{\includegraphics[width=\linewidth]{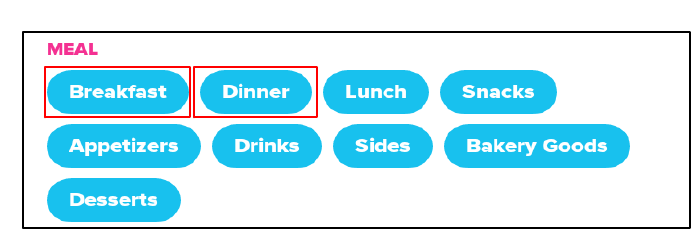}}
    \end{minipage}
    & 
    \begin{minipage}[b]{0.2\columnwidth}
		\centering
		\raisebox{-.5\height}{\includegraphics[width=\linewidth]{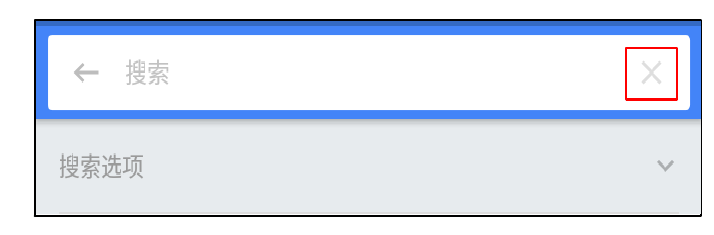}}
    \end{minipage}
    & 
    \begin{minipage}[b]{0.2\columnwidth}
		\centering
		\raisebox{-.5\height}{\includegraphics[width=\linewidth]{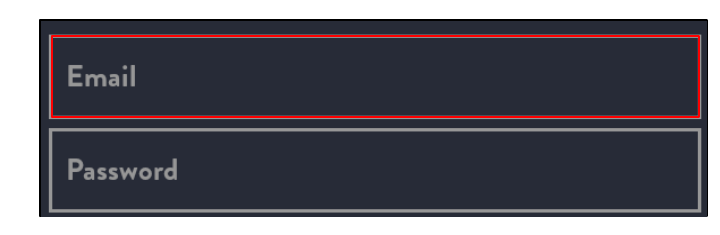}}
    \end{minipage}
    \\ 
    \hline
    \textbf{Ground truth} & Small size & Narrow interval & Low color contrast & Unclear alert information 
    \\
    \textbf{GCN} & Small size & Narrow interval & Low color contrast & Unclear alert information 
    \\
    \textbf{GAT} & Small size & Narrow interval & Low color contrast & Normal \textcolor{red}{$(\times)$} 
    \\
    \textbf{SVM} & Normal \textcolor{red}{$(\times)$} & Normal \textcolor{red}{$(\times)$} & Low color contrast & Normal \textcolor{red}{$(\times)$}
    \\
    \textbf{CNN} & Small size & Normal \textcolor{red}{$(\times)$} & Low color contrast & Normal \textcolor{red}{$(\times)$}
    \\
    \textbf{RF} & Normal \textcolor{red}{$(\times)$} & Small size \textcolor{red}{$(\times)$} & Low color contrast & Normal \textcolor{red}{$(\times)$}
    \\
	 \textbf{MGRL} & Small size & Normal \textcolor{red}{$(\times)$} & Low color contrast & Unclear alert information
	 \\
    \textbf{Mul-RGL} & Small size & Normal \textcolor{red}{$(\times)$} & Low color contrast & Unclear alert information
    \\
    \hline
  \end{tabular}
  \label{tab: models}
\end{table}
As for the RFs model, we experiment with the number of decision trees ranging from 10 to 100 and ultimately trained the model using 20 trees, as it exhibits superior performance at this point.
Notably, in the aforementioned alternative models, the node features and labels are stored separately, and the features are used for training while the labels for model optimization.
For both MGRL and MuL-GRN, they are graph learning models that require as input the GUI-graph we constructed and the feature information of each node within this topology. 
The output of these two models is the classification results of the nodes. 
Additionally, MGRL offers the option to input a perturbed graph structure to test the model's robustness. 
However, in our work, we only need the classification results from the model, so we do not configure this input module.
Afterward, for the second concern, we adopt the global graph attention mechanism~\citep{Wu2022RepresentingLC} to the GAT model, since we need to ensure that the changes of all nodes could be noticed by others.
Notably, for the aforementioned compared works, we provide the same GUI dataset and the same labels to ensure the fairness of comparison.
Also, all these works are executed on the same device as ALVIN and evaluated using identical metrics.
With these concerns well addressed, we then detect the GUIs in our ground truths using these baselines separately.
In reference to~\citep{Chen2020UnblindYA}, we present the predictions of different models detecting the same GUIs and compute their average $Accuracy$ (the number of correct detections is divided by the total number) for detecting our ground truths, respectively.
\\
\textbf{\underline{Results and discussion:}} Four representative examples can be seen in Table~\ref{tab: models} as the qualitative observation of all models' performance.
These four examples are selected to represent the four different types of accessibility issues we can detect.
In general, our method shows the capacity to accurately capture accessibility issues with components in these examples, GAT performs one incorrect detection, CNN detects the accessibility issues correctly in two examples, while SVM and RF correctly annotate only one example.
To be specific, the first example is the issue of small component size, GCN, GAT, CNN, MGRL, and Mul-RGL are able to correctly capture this kind of issue, but SVM and RF classified it as the normal component.
For the second example, it shows the issue of narrow intervals between components.
The model of GCN and GAT can capture this kind of issue correctly, while CNN, SVM, RF, MGRL, and Mul-RGL would not.
In the third example where the color contrast is low, all models can successfully capture such a kind of accessibility issue.
However, for the fourth type of accessibility issue, unclear alert information, our GCN-based method, MGRL, and Mul-RGL could identify this issue, while other models cannot.
Quantitatively, there is a significant difference in the $Accuracy$ of these models after they are utilized to detect our ground truths.
For ALVIN using GCN model, the $Accuracy$ reaches 87.93\%, while for other models, $Accuracy$ (GAT)=63.22\%, $Accuracy$ (SVM)=41.34\%, $Accuracy$ (CNN)=51.28\%, $Accuracy$ (RF)=43.29\%, $Accuracy$ (MGRL)=70.13\%, and $Accuracy$ (Mul-GRN)=72.49\%.

We subsequently analyze the technical reasons for the differences in the performance of each model. 
As for the issue of low color contrast that each model could detect correctly, this is mainly because the attribute related to this kind of issue is only ``intuitive''. 
This attribute directly determines whether the component has such an issue, and the models only need to evaluate the value of this attribute to make successful predictions.
In contrast, issues such as narrow interval involve multiple component attributes that affect decision-making. 
While for other kinds of issues, our proposed GCN method can well detect, but the other models are deficient.
The technical reason is that GCN can fully consider the relationship between each component when training the model, which will affect the results. 
For example, in the navigation bars with similar components, if one component has accessibility issues, the probability that other components connected to it will have accessibility issues increases.
While CNN, SVM, and RF ignore this feature. 
Although the GAT is also able to consider such relations, its attention coefficient magnifies the influence of neighbor nodes and relies only on the edges in graphs~\citep{Velickovic2017GraphAN}.
Therefore, using the GAT, nodes may influence each other even when they are far apart, which would cause inaccurate predictions for normal components due to the long-distance effects.
For instance, if a component located in the top-left of the GUI has an accessibility issue, then the normal components in the bottom-right of this GUI would be affected, causing it might be detected incorrectly.
The incorrect classification of narrow interval by both MGRL and MuL-RGL is primarily due to the handling of node relationships in these two models being highly complex. 
MGRL incorporates additional representation learning for relationships, while MuL-RGL adds multi-level relationship extraction and fusion. 
These mechanisms would highly amplify the impact of relationships between components on the classification results. 
As a result, if one component has an issue or does not have an issue, the connected or indirectly connected components are likely to be assigned the same result.
Meanwhile, the interval between components is significantly influenced by the relationships among them, which makes it challenging for these two models to accurately classify this kind of accessibility issue.
Additionally, these models' over-reliance on the rational setting of hyperparameters, makes it hard to generalize their application to other graph tasks.
Fortunately, these two models are able to classify accessibility issues of unclear alert information that other models could not handle. 
The rationale behind this is that the sensitivity of these two models to features, they are easier to pay attention to such subtle changes in the alert information and capture its features.
Furthermore, compared to ALVIN, both MGRL and MuL-GRN could also avoid detecting invisible redundant information (Challenge 1 in Section~\ref{sec: introduction}), effectively manage minor deviations (Challenge 2), and are scalable (Challenge 4). However, these two works fall short in Challenge 3 mentioned in Section~\ref{sec: introduction}, which involves detecting similar components. 
The primary reason is their overemphasis on inter-component relationships, which can lead to erroneous labeling due to multi-level associations with other components.

We also list other reasons for the lower performance of CNN, SVM, and RF as follows.
As for the CNN model, it is more suitable for processing unstructured information (i.e., images), but it is hard to have an ideal performance effect on structured graphs~\citep{LeCun2010ConvolutionalNA}.
For the SVM, it is more applicable to binary classification problems with small samples, and its performance is not ideal for multi-class classification tasks with large-scale data~\citep{Steinwart2008SupportVM}. 
The decision trees constructed by component vector will appear in many similar situations, and hide the real results, resulting in poor performance of the RF~\citep{Breiman2001RandomF}.
Such a performance is also the main reason why we choose the GCN model to implement ALVIN.

\vspace{-0.42em}
\subsection{Threats to validity}
Any human-involved study can incur threats to validity.
The first threat is that our adopted datasets are not entirely annotated by low vision users, while inviting 10 highly nearsighted volunteers to assist us.
The way these nearsighted users operate the apps and their perceptions are different from those of low vision users, causing us might obtain unfairly labeled datasets.
To best alleviate this threat, we recruit low vision participants with rich experience in using apps, and adopt a joint discussion method to annotate the inaccessible components in GUIs.
While for the highly nearsighted volunteers, we ask low vision users to teach and train these volunteers, and require them to annotate GUIs after their annotations of 10 practicing GUIs are almost identical to those of low vision users.
After this training, these volunteers are allowed to annotate the components in our ground truths.
Although this method cannot guarantee that the annotated results of highly nearsighted users are completely consistent with those of low vision users, we try our best to ensure the fairness of experiments and provide the greatest treatment to low vision users.

The other pivotal risk is that the 500 apps we adopt in this work are not randomly selected, but crawled from the recommended list in Google Play.
This makes ALVIN hard to represent the practical situation of all apps.
We mitigate such a validity by collecting highly downloaded apps that are also normally used by low vision users.
Also, ALVIN is feasible to a certain extent by using the labels of GUIs in these apps to train, validate, and test.

\subsection{Future work}

In this work, we focus on a critical subset of app accessibility for low vision users, solving the problem of capturing the accessibility issues in GUIs.
We believe it would be fruitful for future studies to fine-tune deep-learning solutions for different types of visual impairment, as well as conduct in-depth research on different levels of low vision users. 
As we collect the datasets, we find such failure cases usually appear in the GUIs with \emph{$<$WebView$>$} or \emph{$<$LinkView$>$}.
While neither ALVIN nor other tools could capture the accessibility issues within these views.
The existing tools for dynamically monitoring  GUIs can help researchers to obtain the information in real time, which might be beneficial to deal with \emph{$<$WebView$>$} and \emph{$<$LinkView$>$}.
The ultimate goal of capturing accessibility issues in GUIs is actually to fix them, so an automated method can not only reduce the workload of developers, but also assist visually impaired users in a timely manner.
Besides, apps in iOS will undoubtedly have accessibility issues, affecting low vision users to access information. 
Therefore, we appeal for the related accessibility research across operating systems could be concerned in the future.
This paper calls for future research along this line.

\vspace{-0.75em}
\section{Conclusion}\label{sec: conclusion}
Mobile app accessibility is the process of ensuring that the contents of apps can be accessed, interacted, and used by anyone, regardless of whether they have a cognitive, visual, auditory, or elderly.
The development of apps should not only cater to the market demand, but also take into account the difference in the ability to acquire information for different groups of people.
Although existing tools could capture the accessibility issues of visual impairment in apps, the results of such rule-based methods might result in inaccurate and redundant labels.
This makes developers hard to improve the accessibility of their products following such tools.

In this paper, we propose a GCN-based method, named ALVIN, to capture the accessibility issues in GUIs with the participation of low vision users.
We evaluate the effectiveness of ALVIN on 48 real-world apps, and compare it with other baseline tools. 
The results show, on average, the F1-score of 81.2\%, Precision of 83.5\%, and Recall of 78.9\% for capturing the accessibility issues, which are all higher than other methods. 
Besides, our conducted ablation experiments revealed that filtering out the invisible redundant information within the GUI has a slight impact on the effectiveness of our method.
The usefulness of ALVIN is examined by submitting issues to 20 open-source apps, and it is encouraged that 18 of 20 submitted issues are fixed or under fixing.
Also, compared to other ML-methods, our adopted GCN model outperforms them.

\section{Availability of data and materials}\label{sec: data}
All datasets and scripts involved in this work are open-source in \url{https://doi.org/10.5281/zenodo.11933637}.

\section{Acknowledgment}\label{sec: acknowledgment}
The work is funded by ``the Fundamental Research Funds for the Central Universities,JLU, 2022-JCXK-16'', ``the National Natural Science Foundation of China (NSFC) No. 62102160'', and supported by ``Jilin Provincial Natural Science Foundation, 20230101070JC''.
\bibliographystyle{plain}
\bibliography{accessibility}


\end{document}